\documentclass[twocolumn]{svjour3}          
\smartqed  
\usepackage{pgfpages}
\usepackage{cite}
\usepackage{graphicx}
\usepackage{amsmath,amssymb,amsfonts}
\usepackage{textcomp}
\usepackage{xcolor}
\usepackage{subfigure}
\usepackage{authblk}
\usepackage{appendix}
\usepackage{todonotes}
\usepackage{framed}
\usepackage{multirow}
\usepackage{soul}
\usepackage{algorithm}
\usepackage{algpseudocode}  
\newcommand{\hilite}[1]{\textcolor{black}{#1}}
\journalname{Preprint}
\begin{document}
\sloppy 
\title{SUR-FeatNet: Predicting the Satisfied User Ratio Curve for Image Compression with Deep Feature Learning\thanks{Funded by the Deutsche Forschungsgemeinschaft (DFG, German Research Foundation) -- Project-ID 251654672 -- TRR 161 (Project A05).}
}

\titlerunning{Predicting the SUR Curve for Image Compression}        
\authorrunning{H. Lin et al.}        

\author{Hanhe Lin$^{1}$ \and
Vlad Hosu$^{1}$ \and
Chunling~Fan$^{2}$ \and Yun~Zhang$^{2}$ \and
Yuchen~Mu$^{3}$ \and
Raouf~Hamzaoui$^{4}$ \and Dietmar Saupe$^{1}$}
\institute{
$^1$~Department of Computer and Information Science, University of Konstanz, Germany. 
\email{hanhe.lin@uni-konstanz.de}
\\
$^2$~Shenzhen Institutes of Advanced Technology, Chinese Academy of Sciences, China.\\
$^3$~School of Engineering, University of Edinburgh, UK. \\
$^4$~School of Engineering and Sustainable Development, De Montfort University, UK. \\
}

\date{Received: date / Accepted: date}

\maketitle

\begin{abstract}
 The satisfied user ratio (SUR) curve for a lossy image compression scheme, e.g., JPEG, characterizes the complementary cumulative distribution function of the just noticeable difference (JND), the smallest distortion level that can be perceived by a subject when a reference image is compared to a distorted one. A sequence of JNDs can be defined with a suitable successive choice of reference images. We propose the first deep learning approach to predict SUR curves. We show how to apply maximum likelihood estimation and the \hilite{Anderson-Darling} test to select a suitable parametric model for the distribution function. We then use deep feature learning to predict samples of the SUR curve and apply the method of least squares to fit the parametric model to the predicted samples. Our deep learning approach relies on a siamese convolutional neural network, transfer learning, and deep feature learning, using pairs consisting of a  reference image and a compressed image for training. Experiments on the MCL-JCI dataset showed state-of-the-art performance. For example, the mean Bhattacharyya distances between the predicted and ground truth first, second, and third JND distributions were 0.0810, 0.0702, and 0.0522, respectively, and the corresponding average absolute differences of the peak signal-to-noise ratio at a median of the first JND distribution were 0.58, 0.69, and 0.58 dB. \hilite{Further experiments on the JND-Pano dataset showed that the method transfers well to high resolution panoramic images viewed on head-mounted displays.}
\end{abstract}

\section{Introduction}
Image compression is typically used to meet constraints on transmission bandwidth and storage space. The quality of a compressed image is quantitatively determined by encoding parameters, e.g., the quality factor (QF) in JPEG compression. When images are compressed, artifacts such as blocking and ringing may appear and affect the visual quality experienced by the users.
The satisfied user ratio (SUR) is the fraction of users that do not perceive any distortion when comparing the original image to its compressed version. The constraint on the SUR may vary according to the application. 


Determining the relationship between the encoding parameter and the SUR is a challenging task. The conventional method consists of three steps. First, the source image is compressed multiple times at different bitrates. Next, a group of subjects is asked to identify the smallest distortion level that they can be perceived. 
A subject cannot notice the distortion until it reaches a certain level. This just noticeable difference (JND) level is different from one subject to another due to individual variations in the physiological and visual attention mechanisms.
Finally, the overall SUR for the image is obtained by statistical analysis.  
Following this procedure, several subjective quality studies were conducted and yielded JND-based image and video databases, e.g., MCL-JCI~\cite{MCL-JCI}, \hilite{JND-Pano}~\cite{liu2018jnd}, SIAT-JSSI~\cite{FanCL_JVCIR19}, SIAT-JASI~\cite{FanCL_JVCIR19}, MCL-JCV~\cite{MCL-JCV}, and VideoSet~\cite{Videoset}. 
Subjective visual quality assessment studies are reliable but time-consuming and expensive. In contrast, objective (algorithmic) SUR estimation can work efficiently at no extra annotation cost. 

In recent years, deep learning has made tremendous progress in computer vision tasks such as image classification \cite{szegedy2016rethinking}\cite{he2016deep}, object detection \cite{redmon2016you}\cite{lin2017focal}, and image quality assessment (IQA) \cite{bosse2018deep}\cite{hosu2019koniq}\cite{wiedemann2018disregarding}. Instead of carefully designing handcrafted features, deep learning-based methods automatically discover representations from raw image data that are most suitable for the specific tasks, and can improve the performance significantly. 

Inspired by these findings, we propose a novel deep learning approach to predict the relationship between the SUR and the encoding parameter (or distortion level) for compressed images. Given a pristine image and its distorted versions, we first use a siamese network \cite{bromley1994signature, chopra2005learning} to predict the SUR at each distortion level. Then we apply the least squares method to fit a parametric model to the predicted values and use the graph of this model as SUR curve. 

The main contributions of our work are as follows:
\begin{enumerate}

\item We exploit maximum likelihood estimation (MLE) and the \hilite{Anderson-Darling} test to select the most suitable parametric distribution for SUR modelling instead of using the normal distribution as a default like all previous works.

\item We propose a deep learning architecture to predict samples of the SUR curves of compressed images automatically, followed by a regression step yielding a parametric SUR model.


\item We improve the performance of our model by using transfer feature learning from a similar prediction task. We first train the proposed model independently on an IQA task. Given the images for SUR prediction, we extract multi-level spatially pooled (MLSP) \cite{hosu2019effective} features from the learned model, on which a shallow regression network is further trained to predict the SUR value for a given image pair. 
\end{enumerate}

Compared to our previous work \cite{fan2019net}, our new contributions are as follows. (1) We optimize the proposed architecture and apply feature learning instead of a fine-tuning approach, significantly decreasing computational cost and improving performance. (2) We use MLE and the \hilite{Anderson-Darling} test to select the JND distribution model instead of assuming it to be Gaussian. (3) We conduct more experiments using the MCI-JCL dataset to prove the efficiency of our model, providing results for not only the first JND, but for the second and third JNDs as well. \hilite{(4) We add experiments with the JND-Pano dataset, showing that the method transfers well to high resolution panoramic images viewed on head-mounted displays.}

\section{Definitions}
\label{sec_curve}

We consider a lossy image compression scheme that produces monotonically increasing distortion magnitudes as a function of an encoding parameter. The metric for the distortion magnitude may be the mean squared error, and the encoding parameter is assumed to take only a finite number of values. For example, in JPEG, the encoding parameter is the quality factor $\text{QF} \in \{1,\ldots,100\}$. A value of QF corresponds to the distortion level $n=101-\text{QF}$, where $n=1$ is the smallest and $n=100$ is the largest distortion level. 



\begin{framed}
\noindent {\bf Definition 1 ($k$th JND).}
For a given pristine image $I[0]$, we associate distorted images $I[n], \; n=1,\ldots, N$ corresponding to distortion levels $n=1, \ldots, N$. Let $\text{JND}_0$ be the (trivial) random variable with probability $\mathbb{P}(\text{JND}_0=0) = 1$. The $k$th JND, which we denote by $\text{JND}_k, \; k \geq 1$, is a random variable whose value is the smallest distortion level that can be perceived by an observer when the image $I[\text{JND}_{k-1}]$ is compared to the images $I[n]$, $n > \text{JND}_{k-1}$. 
\end{framed}

For simplicity of notation, the random variable $\text{JND}_k$ will be denoted by JND when there is no risk of confusion.

\begin{framed}
\noindent {\bf Definition 2 ($p$\% $\text{JND}$).}
The $p$\% $\text{JND}$ is the smallest integer in the set $\{1,2,\ldots, N\}$ for which the cumulative distribution function of $\text{JND}$ is greater than or equal to $\frac{p}{100}$. \end{framed}

Samples of the JNDs can be generated iteratively. The original pristine image $I[0]$ serves as the first anchor image. The increasingly distorted images\linebreak $I[n], \; n=1,2, \ldots$, are displayed sequentially together with the anchor image until a distortion can be perceived. This yields a sample of the first JND. This first image with a noticeable distortion then replaces the anchor image, to be compared with the remaining distorted images sequentially until again a noticeable difference is detected, yielding a sample of the second JND, and so on.



The set of random variables $\{\text{JND}_k \; | \; k \geq 1\}$ in Definition 1 is a discrete finite stochastic process. The 
number of (non-trivial) JNDs of the stochastic process depends on the image sequence on hand. It is limited by the smallest number of JNDs that a random observer is able to perceive for the given image sequence. In practical applications, the first JND is the most important one. At the following JNDs the image quality is degraded multiple times from the original which implies that a satisfactory usage of the corresponding images may be very limited. In this paper, we have considered only the first three JNDs, which was also the choice made in \cite{Videoset}.

Definition 1 is intended for sequences of increasingly distorted images, and the prototype application is given by image compression with decreasing bitrate. However, it can also be applied to other media like sequences of video clips or, more generally, to sequences of perceptual stimuli of any kind. Moreover, these sequences need not be sequences with increasing distortion. JNDs may also be useful, for example, to study the effect of parameter-dependent image enhancement methods.


The notion of a sequence of JNDs obtained by the iterative procedure as considered in this paper was introduced to the field of image and video quality assessment only recently \cite{lin2015experimental}. 
In that contribution, an empirical study for five sequences of compressed images and video clips was carried out, with 20 subjects contributing their sequences of JND samples for each set of stimuli. In the followup paper \cite{MCL-JCI}, a larger dataset of 50 source images was introduced, including subjective tests with 30 participants, and providing the dataset MCL-JCI, that we are using for our studies here. Neither of the mentioned contributions gave a formal definition of JNDs. However, the experimental protocols suggest that in these papers the JND random variables were sampled in the spirit of Definition 1.

\hilite{At this point it is important to take note of the common (but slightly different) usage of JNDs in psychophysics. Those JND scales are based on the long standing principle in psychology that equally noticed differences are perceptually equal, unless always or never noticed. 
It is this linear scale of JNDs that has also been used as units of perceptual quality scales for images in \cite{keelan2003iso,redi2010comparing}. For subjective quality assessment, an input image is compared to a quality ruler, consisting of a series of reference images varying in a single attribute (sharpness), with known and fixed quality differences between the samples, given by a certain number of JND units.}


Another application of this JND scale was given in a later paper \cite{Videoset}, where the $k$th JND for $k>1$ was obtained differently from the procedure outlined in Definition 1, namely by using the same anchor image for all observers. This anchor image was chosen as the one corresponding to the 25\% quantile of the previous JND, i.e., the point at which the fraction of observers that cannot perceive a noticeable difference drops below 75\%. 

\hilite{To conclude, let us state that the classical psychophysical JND scale produces a perceptual distance of one JND unit between the reference image and the first JND  from Definition 1, as expected. However, it is not hard to see that for the $k$th JND, $k>1$, the perceptual distance to the reference image according to the common psychophysical JND scale may differ from the expected value, i.e., $k$.}

\begin{framed}
\noindent {\bf Definition 3 (SUR function  and curve).}
The SUR function is the complementary cumulative distribution function (CCDF) of the JND. The graph of this function is called the SUR curve.
\end{framed}

The SUR function, which we denote by $\text{SUR}(\cdot)$, gives the proportion of the sample population for which the JND is greater than a given value. That is, 
$$\text{SUR}(x)= \mathbb{P}(\text{JND} > x).$$ 
Since the range of the JND is discrete (i.e., integers $\{1,2, \ldots, N\}$), the SUR function is a monotonically decreasing step function.

The SUR curve can be used to determine the highest distortion level for which a given proportion of the population is satisfied, in the sense that it cannot perceive a distortion. Formally, we apply the definition of the SUR function and curve also for the second and third JND, although for these cases, an interpretation as a proportion of ``satisfied'' users is not appropriate.
 
\begin{framed}
\noindent {\bf Definition 4 ($p$\% SUR).}
The $p$\% SUR is the largest integer in the set $\{1,2, \ldots, N\}$ for which the SUR function is greater than or equal to $\frac{p}{100}$,
$$
   p\%\; \text{SUR} = \max \{n \in \{1,...,N\}\;|\;\text{SUR}(n) \ge \frac{p}{100}\}.
$$
\end{framed}

\noindent If we set $p = 75$, we obtain the 75\% SUR used in~\cite{predictionSUR}.

\section{Related works}



Existing research on JND can be classified into three main areas: 1. subjective quality assessment studies to collect JND annotations, 2. mathematical modeling of the JND probability distribution and SUR function, and 3. prediction of the probability distribution of the JND and the SUR curve for a given image or video. 

\subsection{Subjective quality assessment}

\hilite{The JND prediction problem has been addressed for various media, including images and videos, and for different types of applied distortions. Existing JND databases have made this possible.}

Jin~\textit{et al.}~\cite{MCL-JCI} conducted subjective quality assessment tests to collect JND samples for JPEG compressed images and built a JND-based image dataset called MCL-JCI. The tests involved 150 participants and 50 source images. With JND samples for a given image collected from 30 subjects, they found that humans can distinguish only a few distortion levels (five to seven). Since subjective tests are time-consuming and expensive, a binary search algorithm was proposed to speed up the annotation procedure. \hilite{The search procedure helps to quickly narrow down the first noticeable difference, resulting in a smaller number of subjective comparisons than the alternative linear search.}

\hilite{Liu~\textit{et al.}~\cite{liu2018jnd} created a JND dataset for panoramic images viewed using a head-mounted display. JPEG compressed versions of 40 source images of resolution 5000 $\times$ 2500 were inspected by at least 25 observers each. An aggressive binary search procedure was used to identify the corresponding first JNDs.}

Wang~\textit{et al.}~\cite{MCL-JCV} conducted subjective tests on JND for compressed videos using H.264/AVC coding. They collected JND samples from 50 subjects, building a JND-based video dataset called MCL-JCV. 

Wang~\textit{et al.}~\cite{Videoset} built a large-scale JND video dataset called VideoSet for 220 5-s source videos in four resolutions (1080p, 720p, 540p, 360p). Distorted versions of the videos were obtained with H.264/AVC compression. To obtain the JND sample from a given subject, they used a modified binary search procedure comparable to the ones adopted in \cite{MCL-JCI} and \cite{MCL-JCV}. For each subject, samples from the first three JNDs were collected.

Fan~\textit{et al.}~\cite{FanCL_JVCIR19} studied the JND of symmetrically and asymmetrically compressed stereoscopic images for JPEG2000 and H.265 intra-coding. They generated two JND-based stereo image datasets, one for symmetric compression and one for asymmetric compression.

\hilite{We are interested in studying a widely-encountered type of distortion, the JPEG compression. This is why we rely on the MCL-JCI and JND-Pano datasets, which offer JND values for JPEG compressed images.}

\subsection{Mathematical modeling of JND and SUR}

\hilite{In previous works, the distribution of JND values has been modeled as a normal distribution} \cite{MCL-JCV, Videoset}\hilite{, some works have studied its skewness and kurtosis, and others modeled it as a Gaussian mixture} \cite{MCL-JCI}.

In~\cite{MCL-JCV} and ~\cite{Videoset}, a normal distribution was used to model the first three JNDs. In~\cite{Videoset}, the  Jarque-Bera test was used to check whether the JND samples have the skewness and kurtosis matching a normal distribution. Almost all videos passed the normality test.

\hilite{In~\cite{MCL-JCI}, the JND samples are classified into three groups (low QF, middle QF, high QF), and it is assumed that the JND distribution for each group} is a Gaussian mixture with a finite number of components. The parameters of the Gaussian mixture model (GMM) are determined with the expectation maximization algorithm. The number of components of the GMM is determined with the Bayesian information criterion. \hilite{However, this methodology is overly complicated, ambiguous in the choice of the three groups, and not justified.} 


In \cite{FanCL_JVCIR19}, the authors assumed that the JND on the QF scale was normally distributed but also noted that an empirical test ($\beta_2$ test~\cite{ITU-R}) found that only 29 of the 50 source images passed the normality test. In Section~\ref{modeling}, we show that other models are more suitable and propose a method to select one, \hilite{without requiring a complex mixture model.}


\subsection{Prediction of JND and SUR}

\hilite{JND studies evaluate the personal (user-specific) JND and accumulate a distribution of JND values over a population of participants. Existing works have proposed to predict various aspects of the JND distribution, such as the mean value of the JND \cite{measurePredictionJND}, the 75\% JND value \cite{Liujnd2020}, or the actual SUR curve as the \mbox{Q-function} of the fitted normal distribution \cite{predictionSUR}.}

Huang~\textit{et al.}~\cite{measurePredictionJND} propose a support vector regression (SVR)-based model to predict the mean value of the JND for HEVC encoded videos. They exploit the masking effect and a spatial-temporal sensitivity map based on spatial, saliency, luminance, and temporal information.

Wang~\textit{et al.}~\cite{predictionSUR} also use SVR to predict the SUR curve. The SVR is fed a feature vector consisting of the concatenation of two feature vectors. The first one is based on the computation of video multi-method assessment fusion (VMAF) \cite{VMAF} quality indices on spatial-temporal segments of the compressed video, while the second one is based on spatial randomness and temporal randomness features that measure the masking effect in the corresponding segments of the source video. For the dataset VideoSet, the average prediction error at the 75\% SUR between the predicted quantization parameter (QP) value and the ground truth QP value was found to be 1.218, 1.273, 1.345, and 1.605 for resolutions 1080p, 720p, 540p, and 360p, respectively. 

\hilite{Zhang~\textit{et al.}~\cite{Zhan20} use Gaussian process regression to model the relationship between the SUR curve and the bitrate for video compression. Three types of features called visual masking features, recompression features, and basic attribute features are used for training and prediction. Visual masking features consist of one spatial and one temporal feature. Recompression features consist of four different bitrates and one variation of the VMAF score over two different bitrates. Basic attribute features are computed from the anchor video and consist of one VMAF score, the frame rate, the resolution, and the bitrate. Experimental results for VideoSet show that the method outperforms the method in \cite{predictionSUR}}.


Hadizadeh~\textit{et al.}~\cite{Hadizadeh2018PDP} build an objective predictor (binary classifier) to determine whether a reference image is perceptually distinguishable from a version contaminated with noise according to a JND model. Given a reference image and its noisy version, they use sparse coding to extract a feature vector and feed it into a multi-layer neural network for the classification. The network is trained on a dataset obtained through subjective experiments with
15 subjects and 999 reference images. The predictor achieves a classification accuracy of about 97\% on this dataset. 

Liu~\textit{et al.}~\cite{Liujnd2020} propose a deep learning technique to predict the JND for image compression. JND prediction is seen as a multi-class classification problem, which is converted into several binary classification problems. The binary classifier is based on deep learning and predicts whether a distorted image is perceptually lossy with respect to a reference. A sliding window technique is used to deal with inconsistencies in the multiple binary classifications. Experimental results for MCL-JCI show that the absolute prediction error of the proposed model is 0.79 dB peak signal-to-noise ratio (PSNR) on average.

\hilite{Our work improves the modeling and prediction of the JND distribution. We use a deep learning approach. For a general introduction to deep learning we recommend the book \cite{goodfellow2016deep}. Unlike Liu~\textit{et al.}~\cite{Liujnd2020}, we formulate the SUR curve prediction problem as a regression problem. We find a better suited distribution type that matches the empirical JND samples and predict the entire SUR curve, not just a statistic.
}

\begin{table*}[t!]
\centering
\caption{Ranking of the distribution models according to negative log-likelihood of MLE and A-D test for the 50 source images of the MCL-JCI dataset \cite{MCL-JCI} and the 40 images of the JND-Pano dataset \cite{liu2018jnd}. \hilite{The models are from Matlab (R2019b) and described in \cite{Johnson1,Johnson2}}: Half-normal (1), Rayleigh (2), Exponential (3), Generalized Extreme Value (4), Generalized Pareto (5), Stable (6), tLocation Scale (7), Birnbaum-Saunders (8), Extreme Value (9), Gamma (10), Logistic (11),	Loglogistic (12), LogNormal (13), Nakagami (14), Normal (15), Poisson (16), Rician (17), Weibull (18). Results for the two other models available in Matlab, beta distribution and Burr distribution, are not included as fitting the JND samples with the distributions was not possible. \hilite{The log-likelihoods and p-values are available in \cite{hanhe2019}.}} 
\begin{tabular}{cccccccccccccccccccc}
MCL-JCI \cite{MCL-JCI} & Models & 1 & 2 & 3 & 4 & 5 & 6 & 7 & 8 & 9 & 10 & 11 & 12 & 13 & 14 & 15 & 16 & 17 & 18 \\ \hline
\multirow{2}{*}{First JND} & log-likelihood & 16 & 14 & 18 & 2 & 13 & 3 & 1 & 4 & 15 & 6 & 11 & 7 & 5 & 8 & 12 & 17 & 10 & 9  \\  
& A-D reject  & 50 & 18 & 50 & 2  & 47 & 1 & 5 & 3 & 18 & 3 & 2 & 3 & 3 & 4 & 4 & 43 & 4 & 4  \\ 
& A-D rank  & 9 & 6 & 9 & 2  & 8 & 1 & 5 & 3 & 6 & 3 & 2 & 3 & 3 & 4 & 4 & 7 & 4 & 4  \\ 
\hline 
\multirow{2}{*}{Second JND} & log-likelihood & 16 & 13 & 17 & 1 & 14 & 5 & 10 & 2 & 15 & 4 & 11 & 6 & 3 & 7 & 12 & 18 & 9 & 8  \\  
& A-D reject  & 43 & 8 & 49 & 0  & 38 & 0 & 0 & 0 & 6 & 0 & 0 & 0 & 0 & 0 & 1 & 32 & 0 & 0  \\ 
& A-D rank  & 7 & 4 & 8 & 1  & 6 & 1 & 1 & 1 & 3 & 1 & 1 & 1 & 1 & 1 & 2 & 5 & 1 & 1  \\ \hline 
\multirow{2}{*}{Third JND} & log-likelihood & 16 & 13 & 17 & 1 & 14 & 5 & 10 & 2 & 15 & 4 & 11 & 6 & 3 & 7 & 12 & 18 & 9 & 8  \\  
& A-D reject  & 41 & 5 & 44 & 0 & 31 & 0 & 0 & 0 & 7 & 0 & 0 & 0 & 0 & 0 & 0 & 22 & 0 &0 \\
& A-D rank  & 6 & 2 & 7 & 1  & 5 & 1 & 1 & 1 & 3 & 1 & 1 & 1 & 1 & 1 & 1 & 4 & 1 & 1  \\ \hline
 \\[1mm]
\hilite{JND-Pano} \cite{liu2018jnd}  & Models & 1 & 2 & 3 & 4 & 5 & 6 & 7 & 8 & 9 & 10 & 11 & 12 & 13 & 14 & 15 & 16 & 17 & 18 \\ \hline
\multirow{2}{*}{\hilite{First JND}} & log-likelihood & 16 & 15 & 17 & 3  & 1 & 4 & 2 & 13 & 14 & 10 & 9 & 11 & 12 & 7 & 8 & 18 & 6 & 5  \\  
& A-D reject  & 36 & 8 & 40 & 2 & 31 & 1 & 1 & 1 & 0 & 0 & 0 & 0 & 0 & 0 & 0 & 37 & 0 &0 \\
& A-D rank  & 6 & 4 & 8 & 3 & 5 & 2 & 2 & 2 & 1 & 1 & 1 & 1 & 1 & 1 & 1 & 7 & 1 &1 \\
\hline 

\end{tabular}
\label{Table:models}
\end{table*}

\section{Modeling the SUR function}\label{modeling}

\begin{figure}[!t]
\centering
\includegraphics[width=0.8\linewidth]{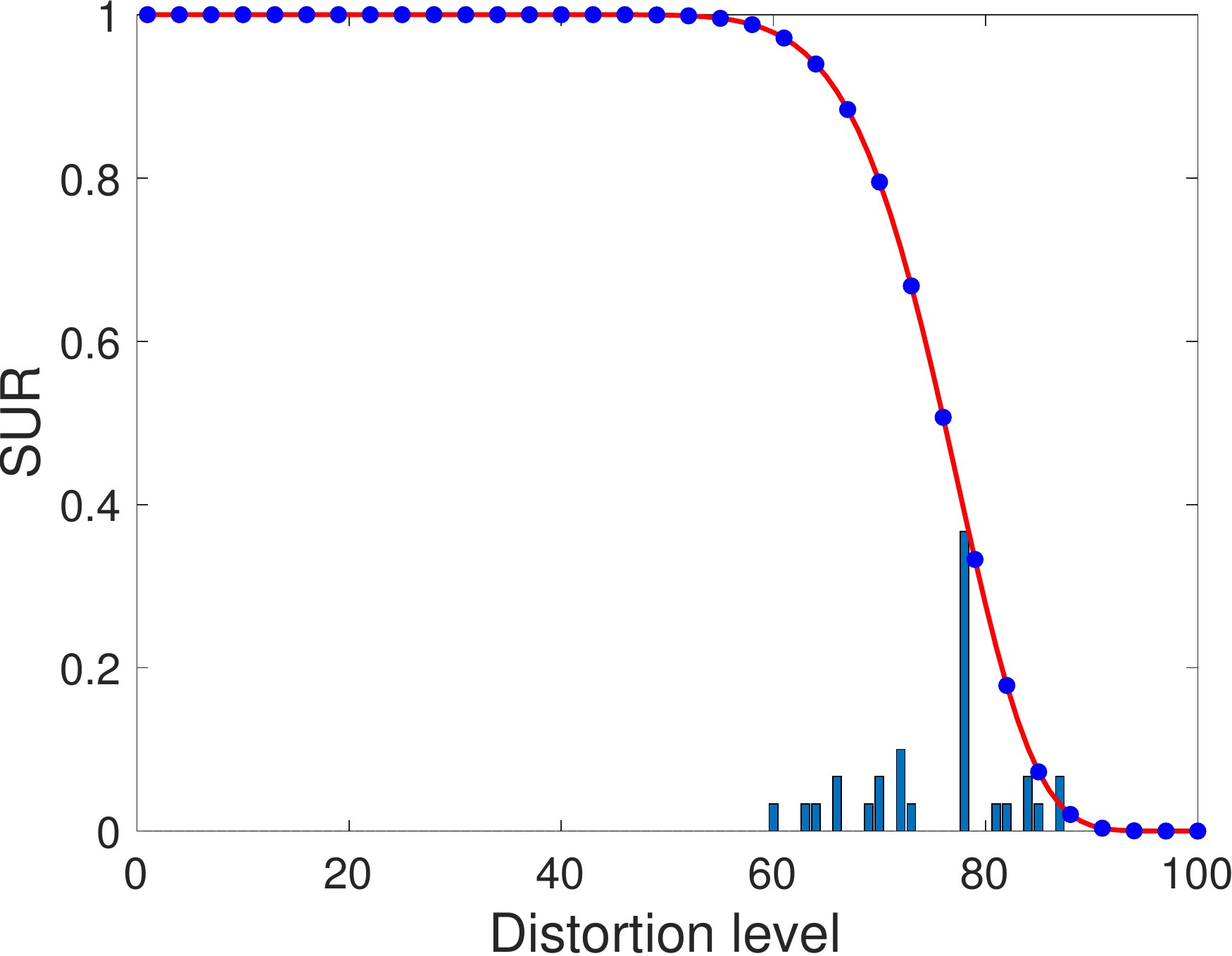}
\caption{\hilite{Illustration of how the ground-truth output values for our prediction model are derived. We start with samples for a JND level from the MCL-JCI  dataset \cite{MCL-JCI}. The histogram, in dark blue, shows their summary. We fit an analytical SUR curve, shown in red, to the empirical samples, given an analytical distribution type. The blue dots show the ground-truth analytical samples that are used to train our prediction model.}}
\label{fig:diag_jnd}
\end{figure}

\hilite{We defined the JND as a discrete random variable and the SUR function as its CCDF, which is a monotonically decreasing step function. In practice, the SUR function must be estimated from sparse and noisy data, i.e., from a small set of subjective JND measurements. We generalize from these samples by fitting a suitable mathematical model to the data. For this purpose, we consider a set of common continuous random variables that have a mathematical form defined by parameters. After choosing the best fitting one, we evaluate the corresponding CCDFs at the integer distortion levels. Thereby, we again obtain a discrete and fitted JND random variable, which replaces the noisy original one for all subsequent steps. The continuous JND distribution provides the ground truth $p$\% JNDs and SURs for the images of the given JND dataset (see Fig.~\ref{fig:diag_jnd} for an illustration of the procedure).}


To select the most suitable distribution for a given dataset of samples, we use maximum likelihood estimation (MLE) and the Anderson-Darling (A-D) test. MLE allows us to estimate the parameters of the probabilistic models and also to rank different models according to increasing  negative log-likelihood, averaged over the source images in the dataset. For a given distribution model and a set of corresponding samples, the A-D test can be applied for the null hypothesis that the JND samples were drawn from the model at a specified significance level (5\% in our experimental settings). This allows us to rank the models according to the number of times the null hypothesis was rejected. \hilite{The A-D test was a suitable goodness of fit test for the datasets considered in this paper. Unlike the chi-squared test, it can be used with a small number of samples. It is also more accurate than the Kolmogorov-Smirnov test when the distribution parameters are estimated from the data \cite{nist}.}  

We considered the 20 parametric continuous probability distribution models that are available in Matlab (R2019b)
and fitted them to the JND samples of the MCL-JCI \cite{MCL-JCI} \hilite{and JND-Pano \cite{liu2018jnd}} datasets, expressed in terms of distortion levels and also in the reverse orientation, i.e., with respect to the corresponding JPEG quality factors QF. 
Considering the two datasets together, the generalized extreme value (GEV) distribution, applied for the QF data, was the most suitable model.


Table \ref{Table:models} shows the results for the QF data. For the 50 source images in the MCL-JCI dataset, the GEV distribution ranked second in terms of both the negative log-likelihood and the A-D test for the first JND. In contrast, the Gaussian distribution, ranked 12th for the negative log-likelihood criterion and 4th for the A-D test. 
\hilite{For the JND-Pano dataset, the GEV distribution ranked third for both the log-likelihood and the A-D test.}

The probability density function (PDF) of the GEV distribution is given by
\begin{equation} \label{eq:gev}
    f(x|\xi,\mu,\sigma) = \frac{1}{\sigma} \exp(-z^{-\frac{1}{\xi}})
z^{-1-\frac{1}{\xi}}
\end{equation}
where $x \in \mathbb{R}$ satisfies
$$
    z = 1+\xi\frac{x-\mu}{\sigma} >0
$$
Here, $\xi \neq 0$, $\mu$, and $\sigma$ are called shape parameter, location parameter, and scale parameter, respectively. 

Since convergence of MLE was better for the QF data than for the distortion level data, we built our models based on the QF data. That is, we used the PDF 
\begin{equation} \label{eq:f_Y}
    f_Y(y) = f_X(101-y\,|\,\xi,\mu,\sigma)
\end{equation}
to model the JND distribution, where $f_X$ is the PDF of the GEV that models the QF data. Note that $f_Y$ is not the PDF of a GEV distribution. Finally, the CCDF of $f_Y$,
$$
   \overline{F}(y\,|\,\xi,\mu,\sigma) = 1 -  \int_{-\infty}^y f_Y(s\,|\,\xi,\mu,\sigma) \, ds,
$$
served as a model for the SUR function, where we have copied the GEV parameters of $f_X$ in the notation of $f_Y$ and $\overline{F}$ for convenience.

\hilite{Finally, to return to a discrete model for the JND, we sample the continuous model $\overline{F}(y\,|\,\xi,\mu,\sigma)$ at integer distortion levels $y=1,\ldots,100$ and arrive at the piecewise constant SUR function
$$
   \text{SUR}(y) = \left\{ 
\begin{array}{ll}
   1  & y < 1 \\
   \overline{F}(\lfloor y \rfloor\,|\,\xi,\mu,\sigma)& y \ge 1
\end{array}
\right.
$$
where $\lfloor y \rfloor$ denotes the greatest integer less than or equal to $y$. For completeness, the modeled JND is given by the discrete random variable
$$
\mathbb{P}(\text{JND}=n) = 
   \text{SUR}(n-1) - \text{SUR}(n) .
$$
}

\begin{figure*}[t]
\centering
\begin{minipage}{0.32\textwidth}
\includegraphics[width=1.0\linewidth]{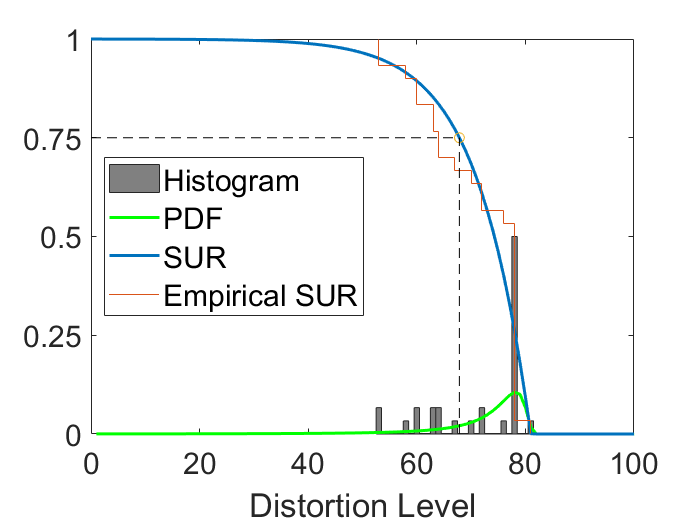}
\centerline{(a) First JND}
\end{minipage} 
\begin{minipage}{0.32\textwidth}
\includegraphics[width=1.0\linewidth]{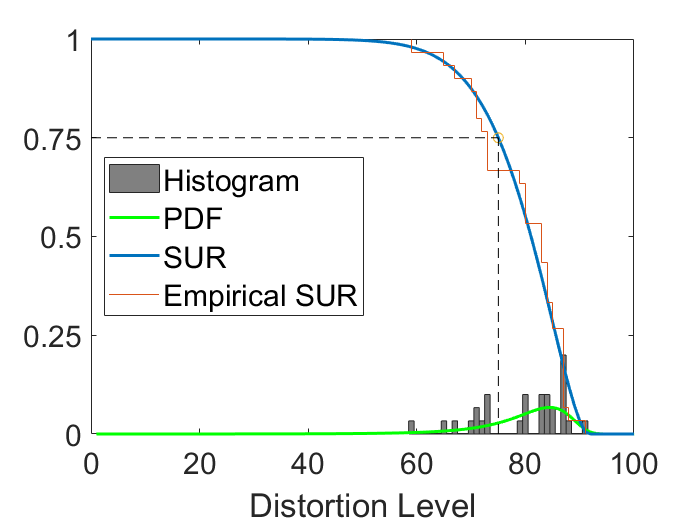}
\centerline{(b) Second JND}
\end{minipage}
\begin{minipage}{0.32\textwidth}
\includegraphics[width=1.0\textwidth]{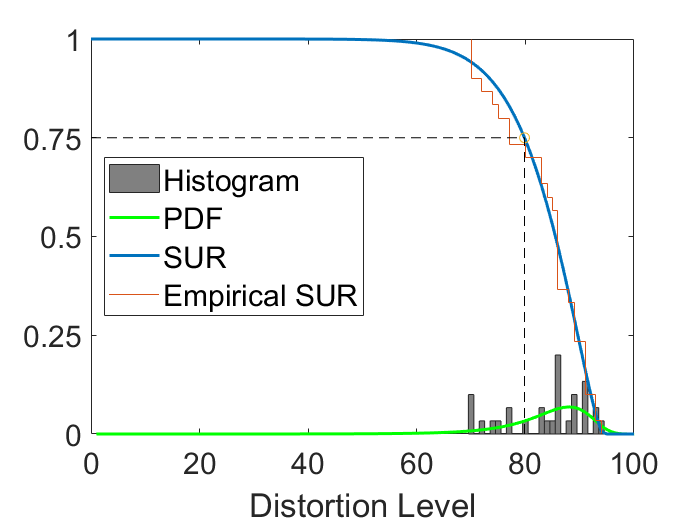}
\centerline{(c) Third JND}
\end{minipage}
\caption{ SUR curve and 75\% SUR of the first three JNDs. The data is for the fifth source image in the MCL-JCI dataset \cite{MCL-JCI}.}
\label{fig:example}
\end{figure*}

\begin{figure*}[t]
\centering
\begin{minipage}{0.32\textwidth}
\includegraphics[width=1.0\linewidth]{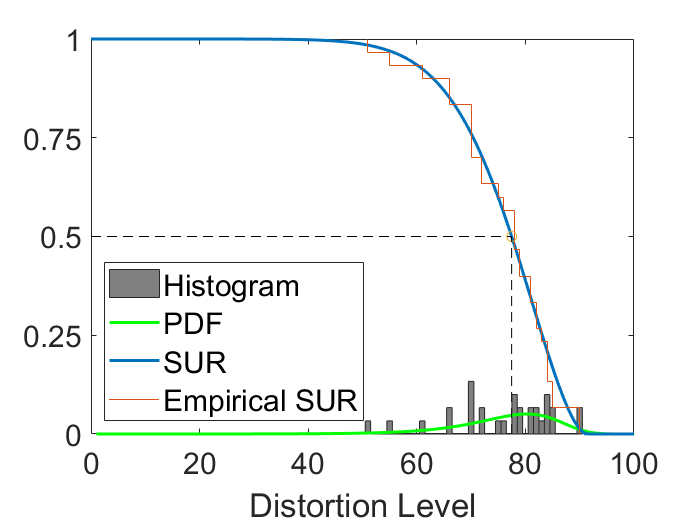}
\centerline{(a) First JND}
\end{minipage} 
\begin{minipage}{0.32\textwidth}
\includegraphics[width=1.0\linewidth]{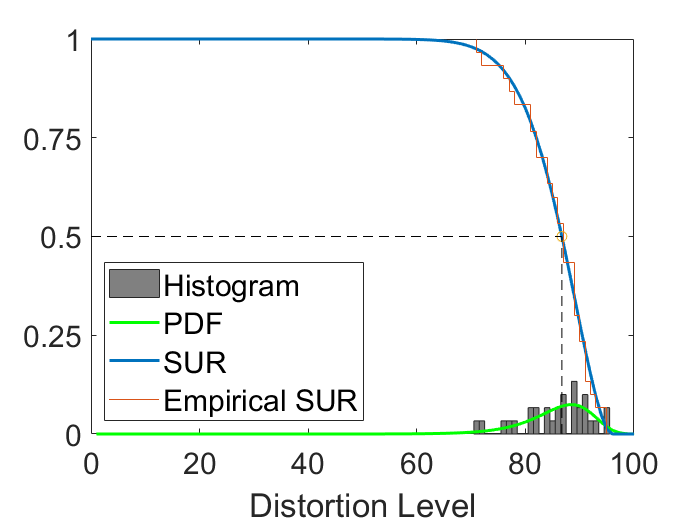}
\centerline{(b) Second JND}
\end{minipage}
\begin{minipage}{0.32\textwidth}
\includegraphics[width=1.0\textwidth]{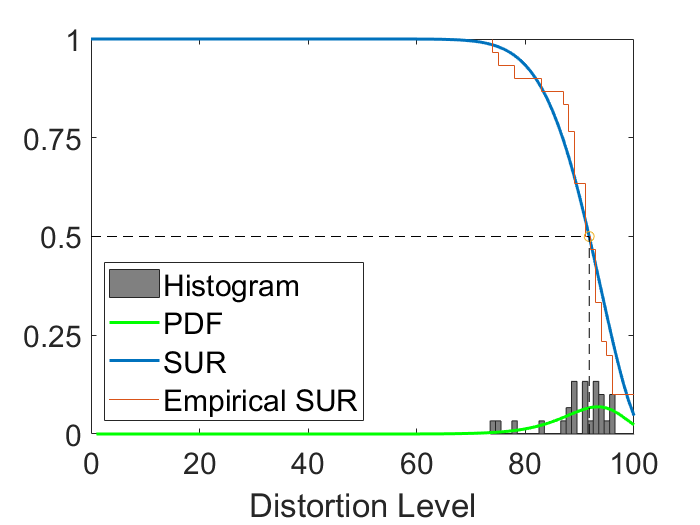}
\centerline{(c) Third JND}
\end{minipage}
\caption{ SUR curve and 50\% SUR of the first three JNDs. The data is for the 14th source image in the MCL-JCI dataset \cite{MCL-JCI}.}
\label{fig:example2}
\end{figure*}




Fig.\ \ref{fig:example}(a) shows the histogram of the first JND for the fifth image in the MCL-JCI dataset, the corresponding empirical SUR curve, the model obtained with MLE of the GEV distribution for the QF data, the corresponding SUR curve, and the 75\% SUR. Fig.\ \ref{fig:example}(b) and Fig.\ \ref{fig:example}(c) show similar results for the second and third JND, respectively. Fig.\ \ref{fig:example2} shows the results for the 14th image in the MCL-JCI dataset, highlighting the 50\% SUR instead of the 75\% SUR. 

\section{Deep learning for SUR prediction}

\subsection{Structure of training data}

We need to predict SUR curves that are calculated from subjective JND studies, given a reference image and a distortion type, e.g., JPEG compression. In order to train a good machine learning model, we considered a few ways to present the available information during training. With respect to the inputs, we could present one (the reference) or more input images (reference and distorted images) to the model. The output has to be a representation of the SUR function. 

With regard to the outputs, for a reference image $I[0]$ and its distorted versions $I[1], \ldots, I[N]$ the SUR curve can be represented as $\text{SUR}(1), \ldots, \text{SUR}(N)$. The $\text{SUR}$ function can be calculated from the empirical CCDF, or by first fitting an appropriate analytical distribution to the subjective data. In the latter case, the analytical representation can be sampled similarly to the empirical SUR or the model can be trained to predict the parameters of the analytical CCDF.

For the inputs of the model, if we attempted to predict a representation of the SUR curve from a single reference image, we would be ignoring information about the particular type of degradation that was applied to images in the subjective study. The model is expected to learn better when both the reference and its distorted version(s) are considered. Ideally we should provide the model with the reference and all the distorted images as inputs. In this way, using an appropriate learning method, the model has all the information that participants in the experiments had, and is expected to perform the best. However, in this formulation the problem is more difficult to solve, \hilite{requiring a different learning model and more training data.} We simplify it by inputting pairs of images: a reference $I[0]$ and a distorted version $I[k]$, $k \in \{1, \ldots, N\}$. In this case we have two options for the outputs: 1. either predict the representation of the entire SUR curve (sampled, or parametric) or 2. predict the corresponding $\text{SUR}(k)$ value. In both cases (1. and 2.), as predictions are independent of each other, the pairwise predictions need to be aggregated into a single SUR curve over all distortion levels for a given reference.


We chose to do pair-based prediction of sampled analytical SUR functions, \hilite{as shown in Fig~\ref{fig:diag_jnd}}. Predicting the empirical samples of the SUR does not perform as well as predicting the sampled analytical SUR. This is probably due to the denoising effect of first mapping a distribution to the subjective data. Each sample of the SUR is independently predicted, and then the overall SUR is estimated from the samples by least-squares fitting.


\subsection{Problem definition}

The regression problem for predicting SUR curves can be formulated as follows. Let $I_1[0], I_2[0], \ldots, I_K[0]$ be a training set of $K$ pristine reference images. For each reference image $I_k[0], \;  k \in \{1,\ldots,K\}$, we associate the $N$ distorted images $I_k[n], \; n=1,\ldots,N$ corresponding to the $N$ distortion levels $n=1, \ldots, N$.


\begin{framed}
\noindent {\bf Problem.}
Let $\text{SUR}_k(\cdot),\;k=1,\ldots,K,$ denote the SUR function of image $I_k[0]$ and its sequence of distorted images $I_k[1],\ldots, I_k[N]$. Find a regression model $f_\theta$, parameterized by $\theta$, such that
$$ 
    f_\theta(I_k[0],I_k[n]) \approx \text{SUR}_k(n)
$$
for $k=1,\ldots,K,~ n = 1,\ldots,N$.
\end{framed}


\subsection{Proposed model}

\begin{figure*}[t]
\centering
\begin{minipage}{0.95\textwidth}
\includegraphics[width=1.0\linewidth]{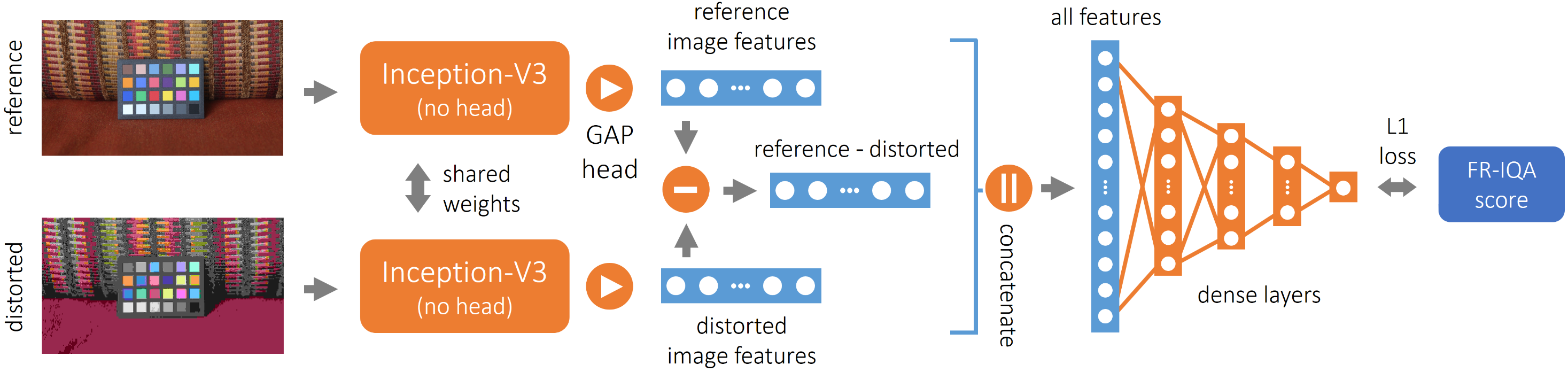}
\centerline{(a)}
\end{minipage}
\begin{minipage}{0.95\textwidth}
\includegraphics[width=1.0\linewidth]{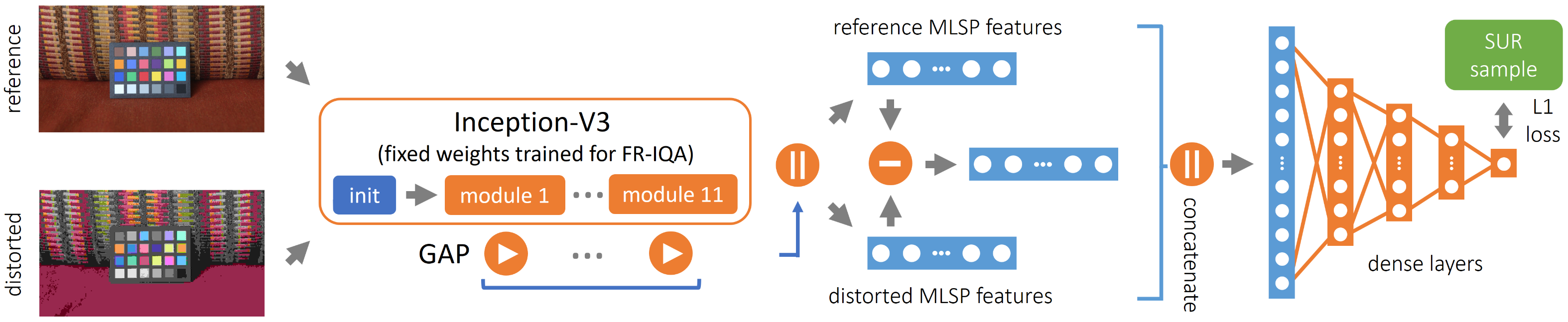}
\centerline{(b)}
\end{minipage}
\caption{SUR-FeatNet architecture for prediction of the SUR curve. In the first stage (a), a Siamese CNN is used to predict an objective quality score of a reference image and its distorted version, which is similar to the SUR prediction task (b) and allowed us to train on a large-scale dataset to address overfitting. In the second stage (b), MLSP features of a reference image and its distorted version were extracted and fed into a shallow regression network that was used to predict SUR values.}
\label{fig:network}
\end{figure*}

Subjective studies are usually time-consuming and expensive, which limits JND datasets to relatively small size. With such small data, training a deep model from scratch may be prone to overfitting. To address this limitation we propose a two-stage model that applies transfer learning and feature learning, as depicted in Fig.~\ref{fig:network}. 

In the first stage (Fig.~\ref{fig:network}a), a pair of images, namely a pristine image $I_k[0]$ and a distorted version $I_k[n]$, are fed into a siamese network that uses an Inception-V3 \cite{szegedy2016rethinking} convolutional neural network (CNN) body with shared weights. The network body is truncated, such that the global average pooling (GAP) layer and the final fully-connected layer are removed. Each branch of the siamese network yields a stack of 2,048 feature maps. The feature maps are passed through a GAP layer, which outputs a 2,048-dimensional feature vector $\mathbf{f}_{\text{gap}}$ for each branch. Then we calculate $\Delta\mathbf{f}_{\text{gap}}$, corresponding to feature vector differences between the distorted images $I_k[n]$ and the pristine image $I_k[0]$, i.e., $$\Delta\mathbf{f}_{\text{gap}} = \mathbf{f}_{\text{gap}}(I_k[0])-\mathbf{f}_{\text{gap}}(I_k[n]).$$
By concatenating the two feature vectors $\mathbf{f}_{\text{gap}}(I_k[0])$, $\mathbf{f}_{\text{gap}}(I_k[n])$ and the feature difference vector $\Delta\mathbf{f}_{\text{gap}}$, we obtain a 6,144 dimensional vector. The latter is passed to three fully connected (FC) layers with 512, 256, and 128 neurons, respectively, where each FC layer is followed by a dropout layer (0.25 ratio) to avoid overfitting. The output layer is linear with one neuron to predict a quality score of the distorted image $I_k[n]$, obtained from a fixed full-reference (FR)-IQA method.

In the second stage (Fig.~\ref{fig:network}b), we keep the weights fixed in the Inception-V3 body as trained in the first stage. A reference and a distorted image are presented to the Inception-V3 body, and for each of them MLSP \cite{hosu2019effective} features  $\mathbf{f}_{\text{mlsp}}$ 
with 10,048 components each are extracted. As in the first stage, we concatenate $\mathbf{f}_{\text{mlsp}}(I_k[0])$, $\mathbf{f}_{\text{mlsp}}(I_k[n])$, and $\Delta\mathbf{f}_{\text{mlsp}} = \mathbf{f}_{\text{mlsp}}(I_k[0])-\mathbf{f}_{\text{mlsp}}(I_k[n])$. The concatenated 30,144-dimensional feature vector is passed to an FC head to predict the SUR value. This FC head has the same structure as the FC head in the first stage. 


Let $(I_r, I_d, q)$ be an item of the training data, where $I_r$ and $I_d$ are the reference image and its distorted version, and $q$ corresponds to the FR-IQA score in the first stage and the SUR value in the second stage. Our objective is to minimize the mean of the absolute error, or L1 loss function
$$
     L =\left|f_\theta(I_r,I_d) - q\right|.
$$

Our proposed model, called SUR-FeatNet, has the following properties. We first train a deep model to predict the FR-IQA score of a distorted image relative to its pristine original. This is similar to  predicting an SUR value and therefore the features learned in the first stage are expected to be useful for predicting SUR values in the second stage. As it is very convenient to generate distorted images given a large-scale set of pristine reference images and to estimate their quality score by an FR-IQA method, training a deep model on a large-scale image set to address overfitting becomes feasible. 

Second, training on these ``locked-in" MLSP features in the second stage instead of fine-tuning a very large deep network not only reduces computational time, but also prevents forgetting previously trained information, which may lead to a better performance on a small dataset.

\subsection{Prediction of the SUR curve and the JND }\label{Sec:fitting}

For any source image $I[0]$,  together with its distorted versions $I[1], \ldots, I[N]$, a sequence of predicted satisfied user ratios  $\text{SUR}(1), \ldots, \text{SUR}(N)$ is obtained from the network. 
Assuming that the JND of the QF data follows a GEV distribution, we estimate the shape parameter $\xi$, the location parameter $\mu$, and the scale parameter $\sigma$ by 
least squares fitting, 
$$
  (\hat{\xi},\hat{\mu},\hat{\sigma})= \arg \min_{\xi,\mu,\sigma} \sum_{n=1}^{N}\left|\overline{F}(n\,|\,\xi,\mu,\sigma)-\text{SUR}(n)\right|^2.
$$
The fitted SUR curve is given by $\overline{F}(n\,|\,\hat{\xi},\hat{\mu},\hat{\sigma})$.

\section{Experiment}

\subsection{Setup}

In our experiments, we used the MCL-JCI dataset \cite{MCL-JCI} and the JND-Pano dataset \cite{liu2018jnd} to evaluate the performance of the proposed method. The MCL-JCI dataset contains 50 pristine images with a resolution of $1920 \times 1080$. Each pristine image was encoded 100 times by a JPEG encoder with QF decreasing from 100 to 1, corresponding to distortion levels 1 to 100. Thus, there are 5,050 images in total. 
\hilite{The JND-Pano dataset contains 40 pristine panoramic images with a resolution of $5000\times 2500$. As for MCL-JCI, each pristine image was encoded 100 times by a JPEG encoder, which resulted in 4,040 images in total.}

The annotation provided for the image sequences in MCL-JCI and for each of the $M = 30$ participants of the study~\cite{MCL-JCI} is the QF value corresponding to the first JND (and also those of the second, third, etc.). 
For each source image $I_k[0]~(k=1, \ldots, 50)$ in the MCL-JCI dataset, 
we modeled its SUR function for the given JND samples, according to the GEV distribution (Eq.\ \eqref{eq:gev}). 
Finally, we sampled the fitted SUR model to derive the target values $\text{SUR}_k(n), k=1,\ldots,50, n=1,\ldots,100$ for the deep learning algorithm. \hilite{Following the same procedure, we derived the target values $\text{SUR}_k(n), k=1,\ldots,40, n=1,\ldots,100$ of the first JND in the JND-Pano dataset, which contains 19 to 21 JND measurements per image (after outlier removal).}

\begin{table*}[!t]
\centering
\begin{tabular}{l  c  c  c }
& Bhattacharyya & $\Delta \text{JND}$  & $\Delta \text{PSNR}$ \\
Scheme &distance &    & (dB) \\ \hline
1. Fine-tune (ImageNet) &0.1244 &5.86&0.74 \\
2. Fine-tune (KADIS-700k) &0.0936 & 5.17 & 0.64 \\
3. MLSP (ImageNet) & 0.0949 & 5.22 & 0.64 \\
4. MLSP (KADIS-700k) &\bf 0.0810 &\bf 4.44 &\bf 0.58
\end{tabular}
\caption{Performance comparison for the first JND of MCL-JCI with different learning schemes. $\Delta \text{JND}$ is the MAE of the 50\% JNDs. $\Delta \text{PSNR}$ is the MAE of the PSNR at the 50\% JNDs.}
\label{tb:perf_comp}
\end{table*}

$k$-fold cross validation was used to evaluate the performance ($k=10$). Specifically, each dataset was divided into 10 subsets, each containing a certain number of source images (five images in MCL-JCI and four images in JND-Pano) and all corresponding distorted versions of them. Each time, one subset was kept as a test set, and the remaining nine subsets were used for training and validation. The overall result was the average of 10 test results. 

The Adam optimizer \cite{kingma2014adam} was used to train SUR-FeatNet with the default parameters $\beta_1 = 0.9$, $\beta_2 = 0.999$, and a custom learning rate $\alpha$. \hilite{In our experiments, we tried $\alpha = 10^{-1},10^{-2},\dots,10^{-5}$ and found that $\alpha = 10^{-5}$ gave the smallest validation loss. Therefore, we set $\alpha = 10^{-5}$} and trained for 30 epochs. In the training process, we monitored the absolute error loss on the validation set and saved the best performing model. Our implementation used the Python Keras library with Tensorflow as a backend \cite{chollet2015keras} 
and ran on two NVIDIA Titan Xp GPUs, where the batch size was set to $16$. \hilite{The source code for our model is available on GitHub \cite{hanhe2019}}. 

\subsection{Strategies to address overfitting}


For the first stage of our model, we used the Konstanz artificially distorted image quality set (KADIS-700k) \cite{lin2019kadid}. This dataset has 140,000 pristine images, with five degraded versions each, where the distortions were chosen randomly out of a set of 25 distortion types. We used a full-reference IQA metric to compute the objective quality scores for all pairs. For this purpose, we chose MDSI \cite{nafchi2016mean} as it was reported as the best FR-IQA metric when evaluating on multiple benchmark IQA databases. As KADIS-700k is a large-scale set, we only trained for five epochs before MLSP feature extraction.

In addition to transfer learning in the  first stage, we applied image augmentation in the second stage to help avoid overfitting. \hilite{Each original and compressed image of both datasets was split into four non-overlapping patches,} where each patch has a resolution of 960 $\times$ 540 in MCL-JCI and 2500 $\times$ 1250 in JND-Pano. We also cropped one patch of the same resolution from the center of the image. The SUR values for the patches were set to be equal to those of their source images. With this image augmentation, we had 25,250 annotated patches in MCL-JCI and 20,200 annotated patches in JND-Pano.

After training the networks with these training sets, SUR values were predicted for the test set. To predict the SUR of a distorted image, predictions for its five corresponding patches were generated by the network and averaged. 

\begin{figure*}[!htb]
\centering
\begin{minipage}{0.4\textwidth}
\includegraphics[width=1.0\linewidth]{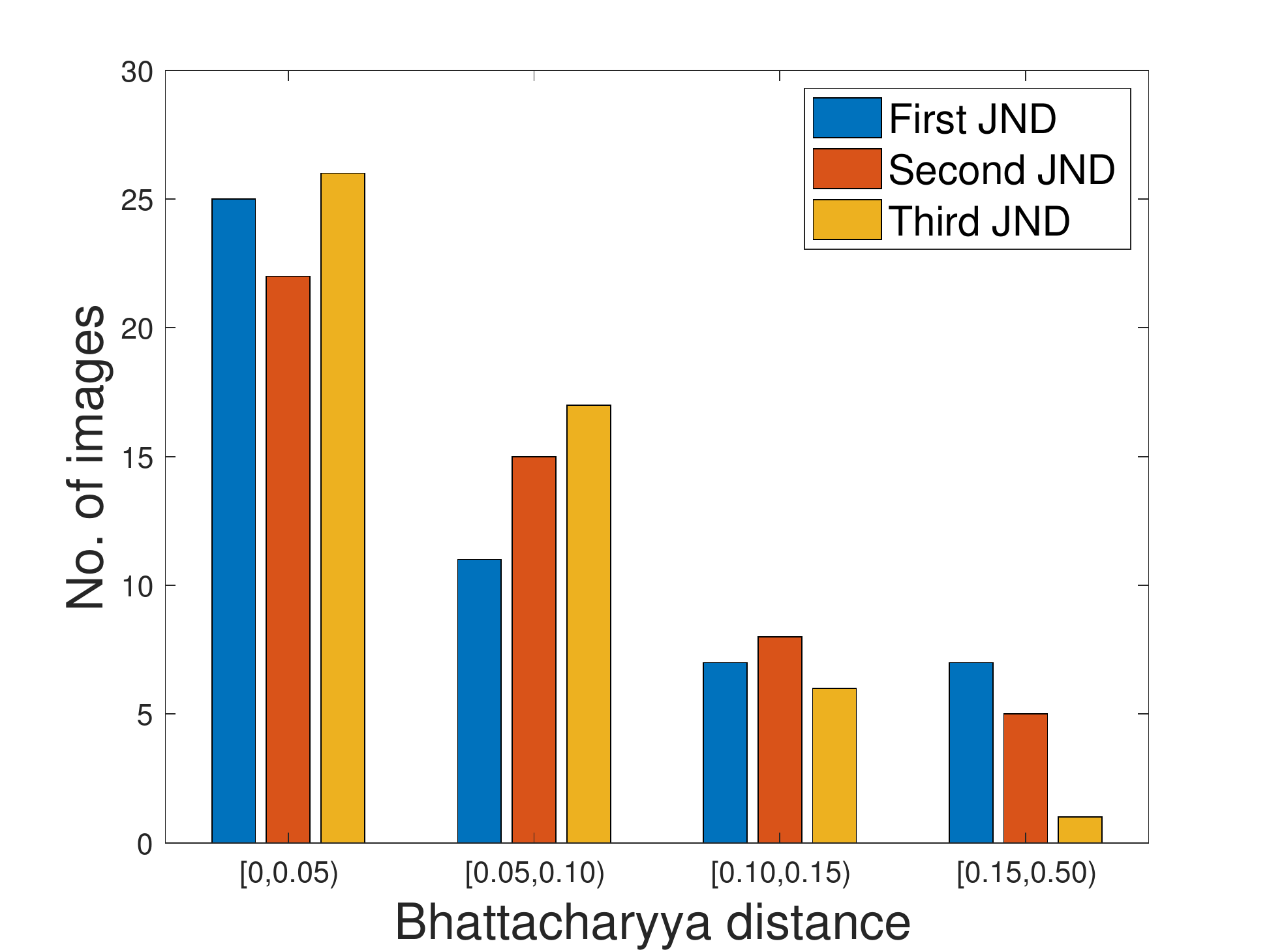}
\centerline{(a)}
\end{minipage}
\begin{minipage}{0.4\textwidth}
\includegraphics[width=1.0\textwidth]{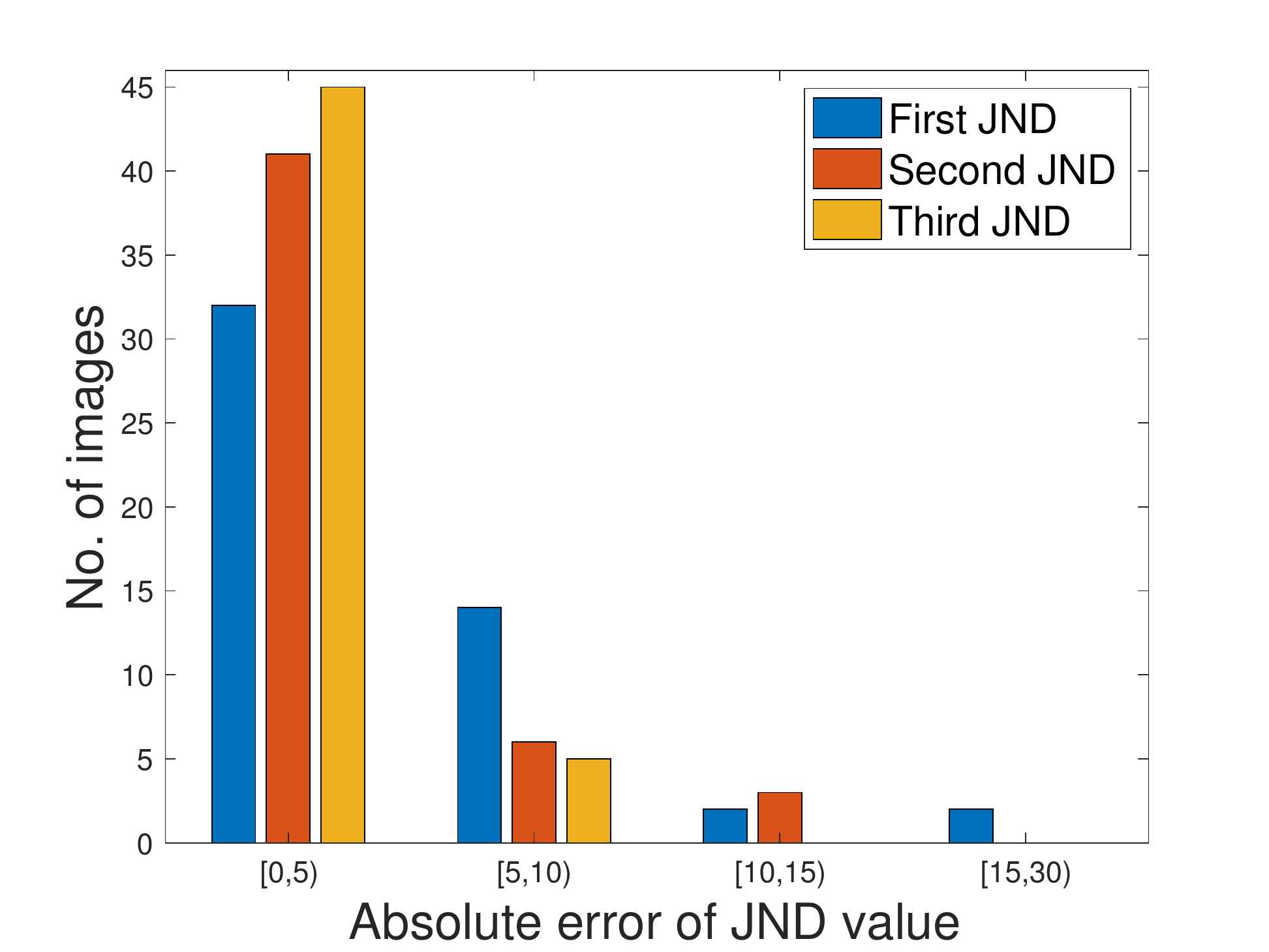}
\centerline{(b)}
\end{minipage}
\caption{Statistics of experimental results on the MCL-JCI dataset. (a) Histogram of Bhattacharyya distance between the predicted JND distribution and the ground truth JND distribution. (b) Histogram of the absolute error between predicted JND (50\% JND) and ground truth JND (50\% JND). The GEV distribution is used as distribution model.}
\label{fig:expresult}
\end{figure*}

\begin{figure*}[!htb]
\centering
\begin{minipage}{0.32\textwidth}
\includegraphics[width=1.0\linewidth]{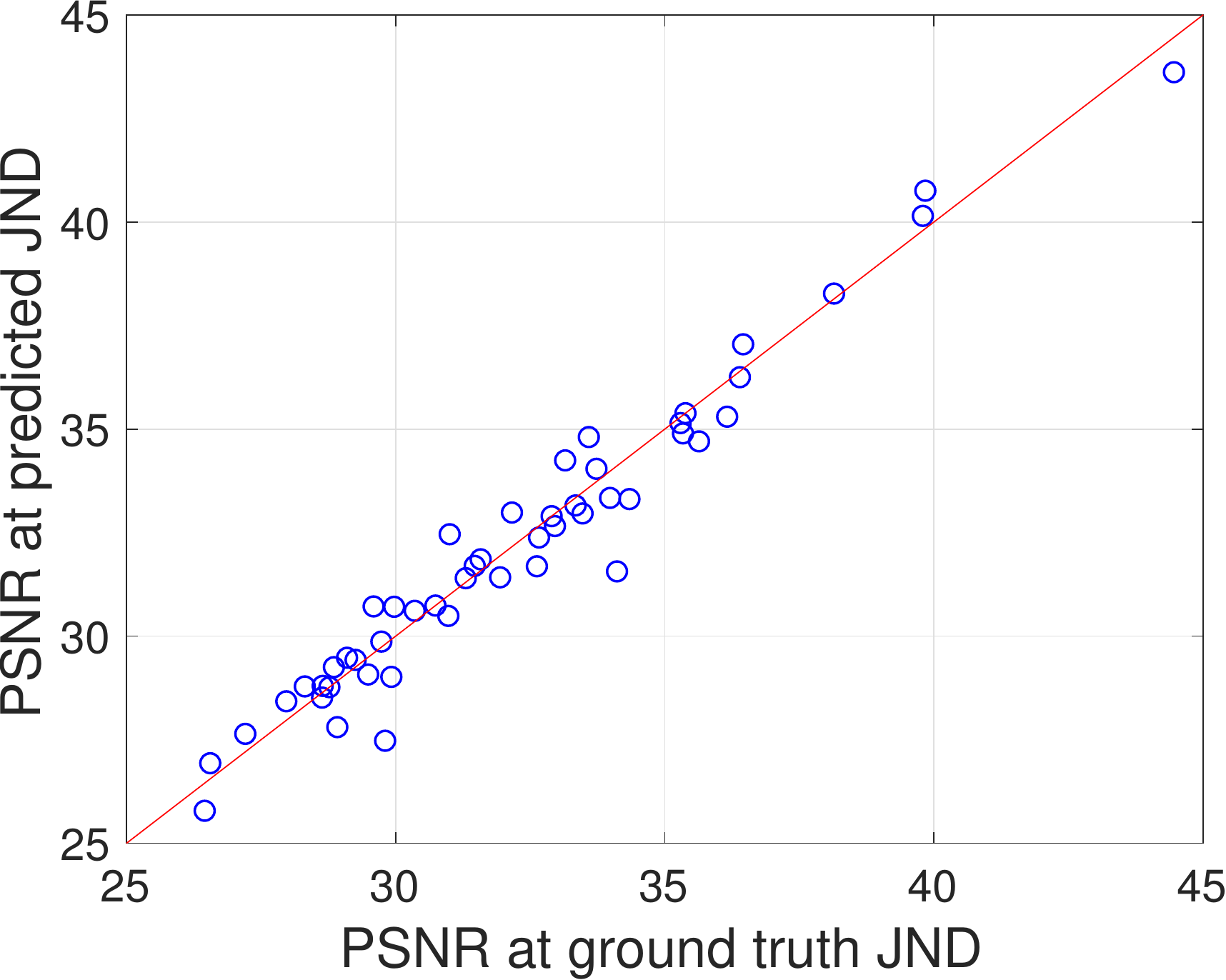}
\centerline{(a) First JND}
\end{minipage} 
\begin{minipage}{0.32\textwidth}
\includegraphics[width=1.0\linewidth]{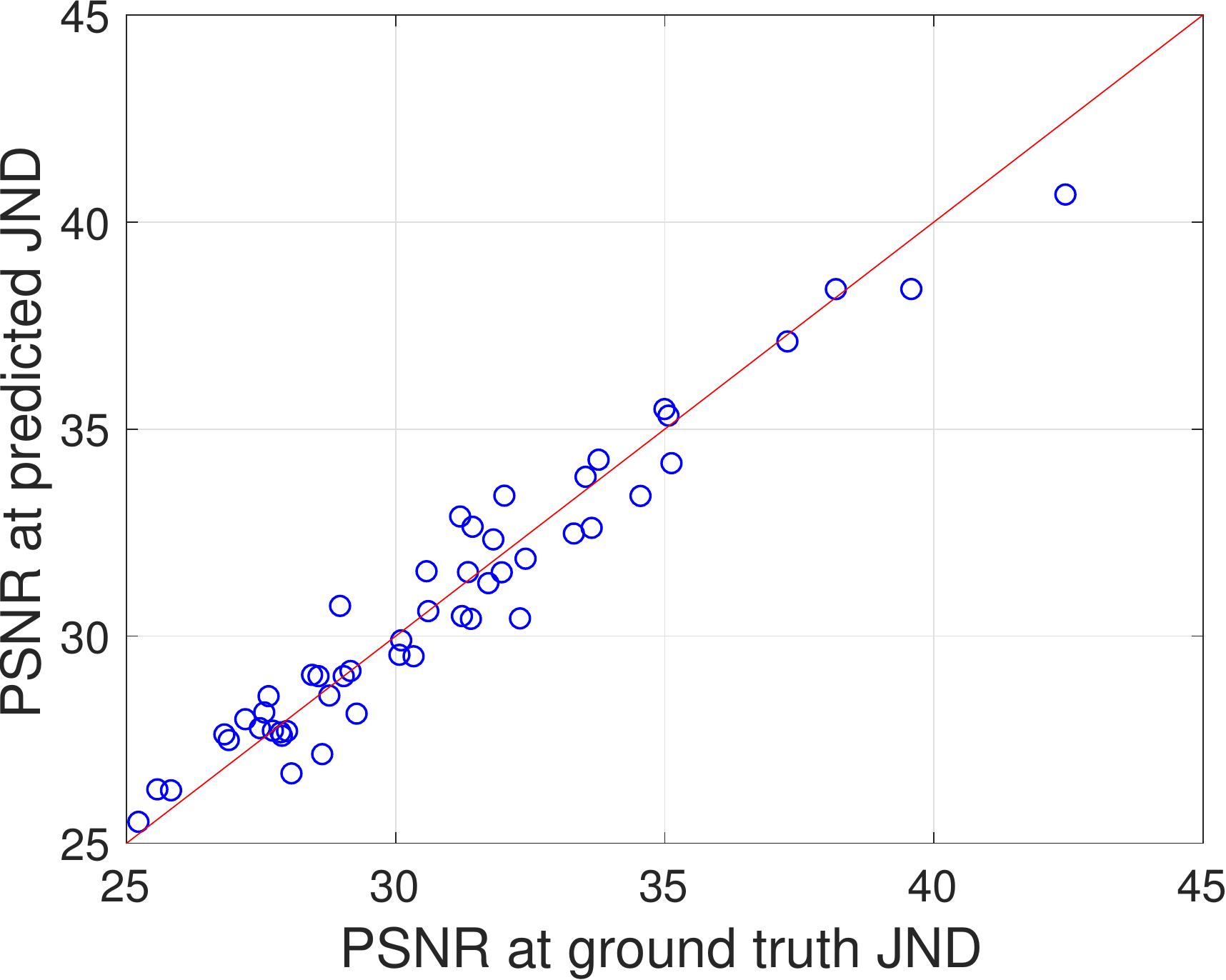}
\centerline{(b) Second JND}
\end{minipage}
\begin{minipage}{0.32\textwidth}
\includegraphics[width=1.0\textwidth]{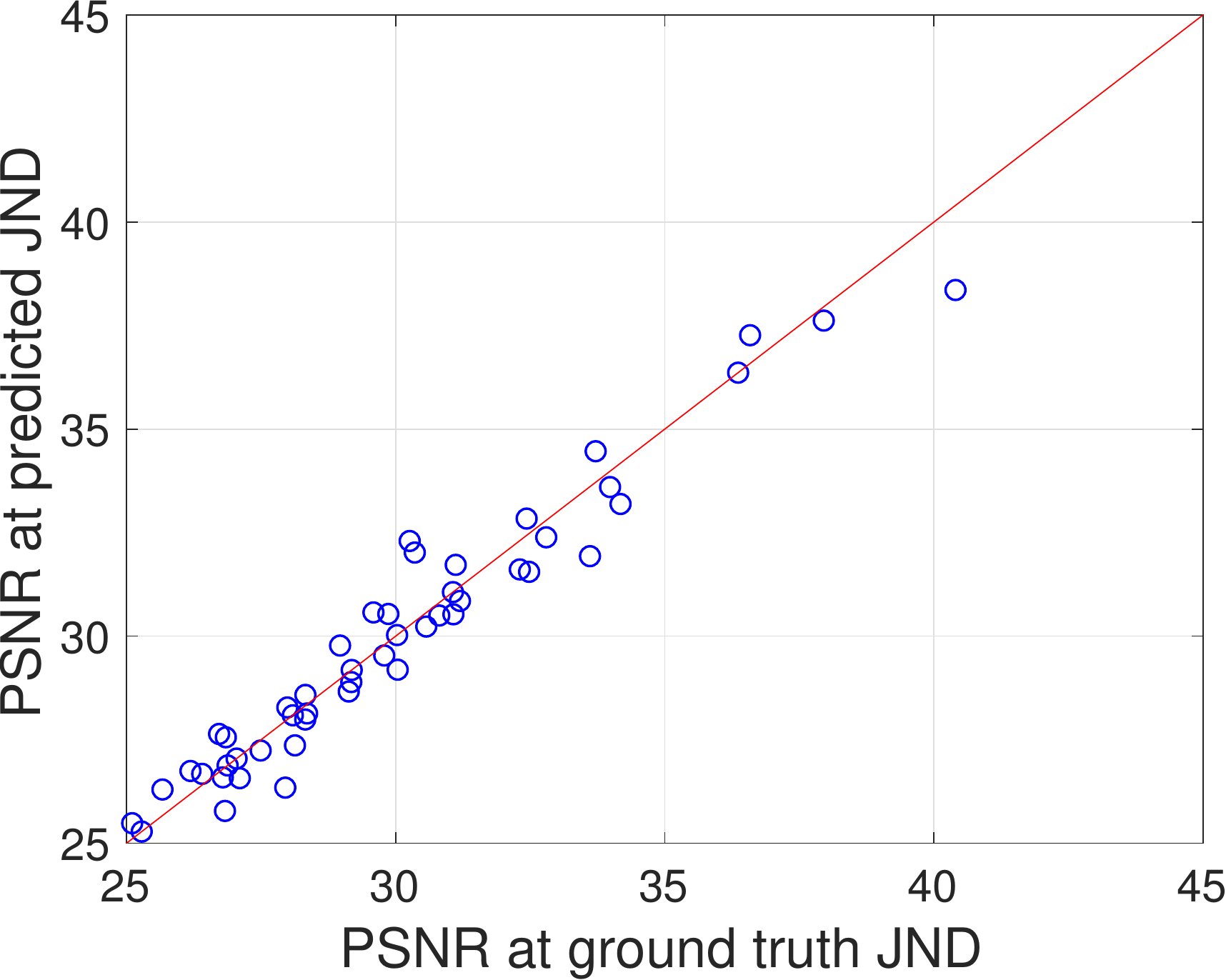}
\centerline{(c) Third JND}
\end{minipage}
\caption{ PSNR comparison between the ground
truth JNDs and predicted JNDs at first JND (a), second JND (b), and third JND (c) for the MCL-JCI dataset. The corresponding PLCCs are 0.9771, 0.9721, and 0.9741.}
\label{fig:psnr_comp}
\end{figure*}

\subsection{Results and analysis for MCL-JCI}


Three metrics were used to evaluate the performance of SUR-FeatNet: mean absolute error (MAE) of the 50\% JNDs, MAE of the PSNR at the 50\% JNDs, and Bhattacharyya distance~\cite{Bhattacharyya1943On} between the predicted and ground truth JND distributions of type GEV. The ground truth GEV parameters were obtained by using MLE to fit a GEV distribution to the MCL-JCI QF values.

We first compared the performance of the following four learning schemes. 

\begin{itemize}
\item[(1)] \textit{Fine-tune (ImageNet).}
In the first scheme, we used the architecture in the first stage (Fig.~\ref{fig:network}a). Its CNN body was initialized with the pre-trained weights on ImageNet and FC layers were initialized with random weights. With the initialized weights, the network was fine-tuned for SUR prediction using the MCL-JCI dataset.
    \item[(2)] \textit{Fine-tune (KADIS-700k).}
The second scheme used the same architecture and same initialized weights as the first scheme. However, it was first fine-tuned on KADIS-700k to predict FR-IQA quality scores before it was fine-tuned on MCL-JCI dataset.
\item[(3)] \textit{MLSP (ImageNet).} In the third scheme, we trained a shallow regression network for SUR prediction based on MLSP features, which were extracted from a pre-trained network on ImageNet.
\item[(4)] \textit{MLSP (KADIS-700k)} The fourth scheme, which is used by our approach, trained the same regression network as the third scheme. However, its MLSP features were extracted from fine-tuned weights on KADIS-700k instead of ImageNet.
\end{itemize}

    Table~\ref{tb:perf_comp} shows the performance of the four schemes for the first JND. Clearly, transfer learning from the image classification domain (ImageNet) to the quality assessment domain (KADIS-700k), together with MLSP feature learning, outperformed the remaining schemes.



Tables~\ref{Table:first_jnd_result}, \ref{Table:second_jnd_result}, and \ref{Table:third_jnd_result} present the detailed results of the first, second, and third JND for each image sequence. 
Fig.~\ref{fig:expresult} shows the statistics. 
For all three JNDs, more than 75\% of the images have a Bhattacharyya distance smaller than 0.1 (Fig.~\ref{fig:expresult}(a)). With respect to the first, second, and third JND, the absolute error in 50\% JND was less than 5 for 32, 41, and 45 images, respectively (Fig.~\ref{fig:expresult}(b)). For more than 90\% of the images, the absolute error in 50\% JND was smaller than 10. Fig.~\ref{fig:psnr_comp} compares the PSNR at ground truth and predicted 50\% JND for the first, second, and third JNDs. The Pearson linear correlation coefficient (PLCC) was very high, reaching 0.9771, 0.9721, and 0.9741, respectively.

\begin{figure*}[h]
    \centering
    \begin{minipage}[c]{0.40\textwidth}
        \includegraphics[width=1.0\linewidth]{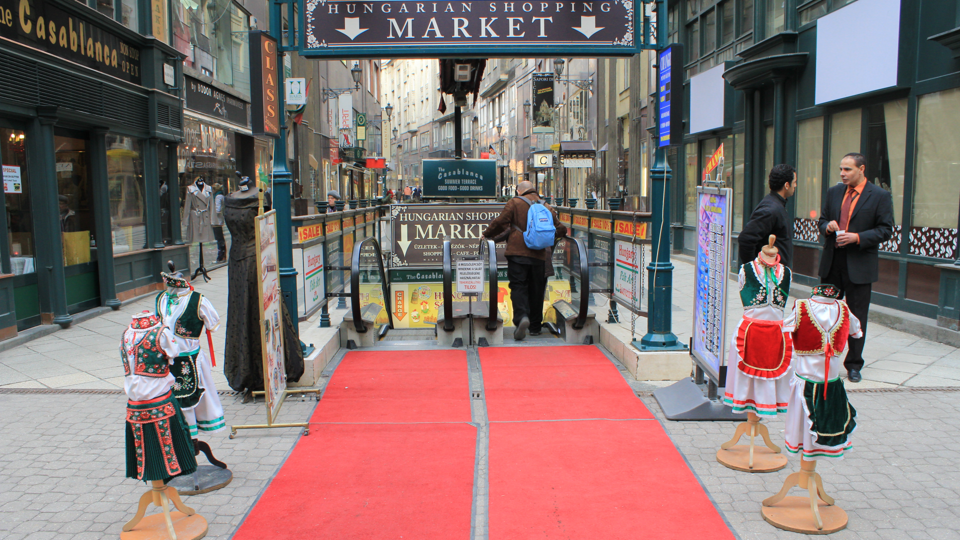}
        \centerline{Image 35}
    \end{minipage}
    \begin{minipage}[c]{0.40\textwidth}
        \includegraphics[width=1.0\linewidth]{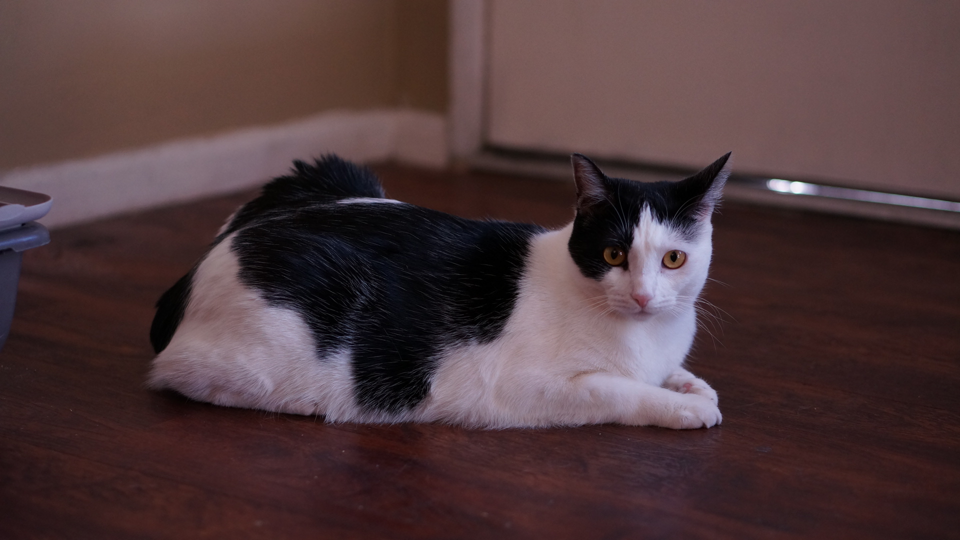}
        \centerline{Image 40}
    \end{minipage}
    \begin{minipage}[c]{0.40\textwidth}
        \includegraphics[width=1\linewidth]{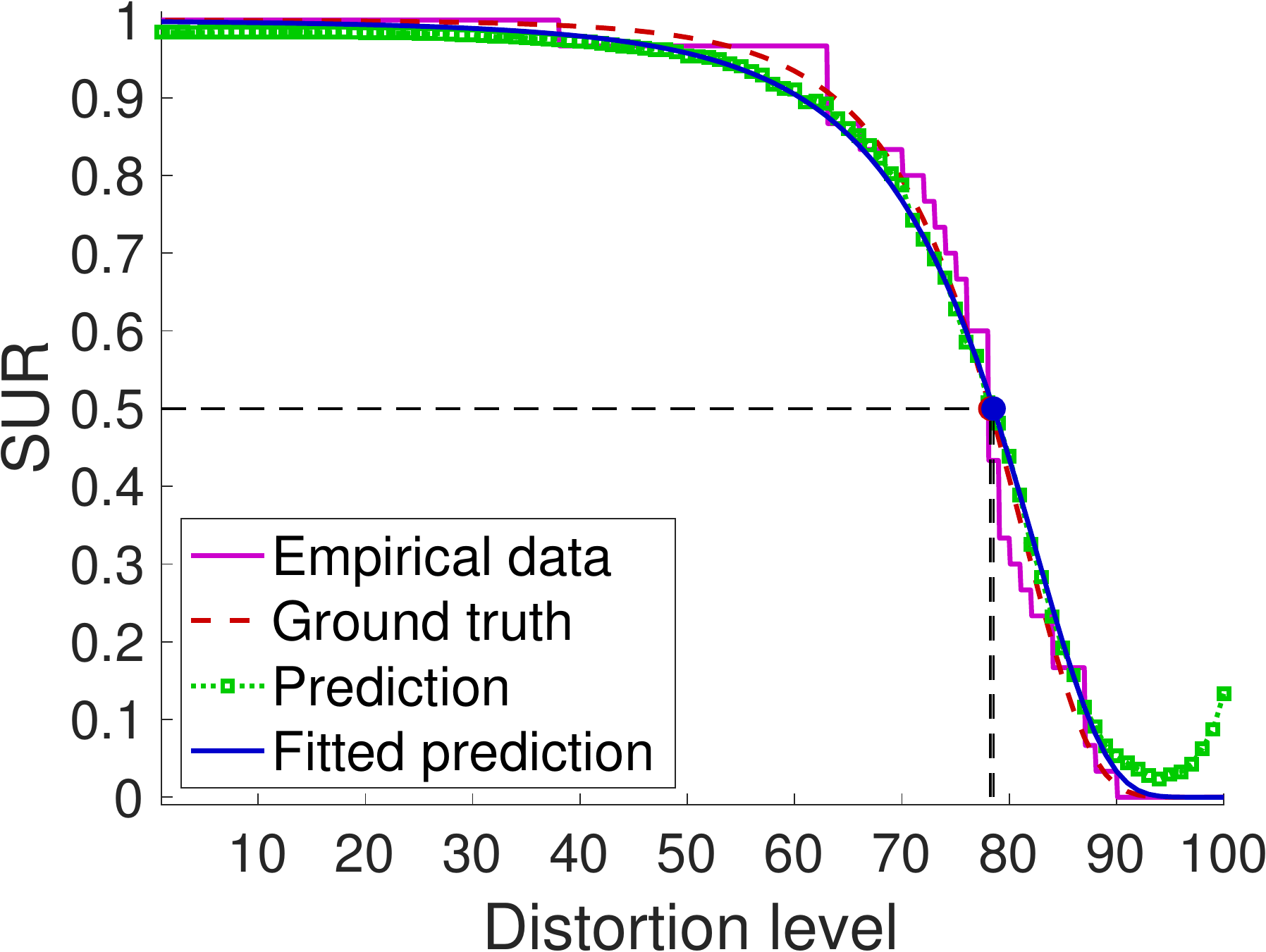}
    \end{minipage}
    \begin{minipage}[c]{0.40\textwidth}
        \includegraphics[width=1\linewidth]{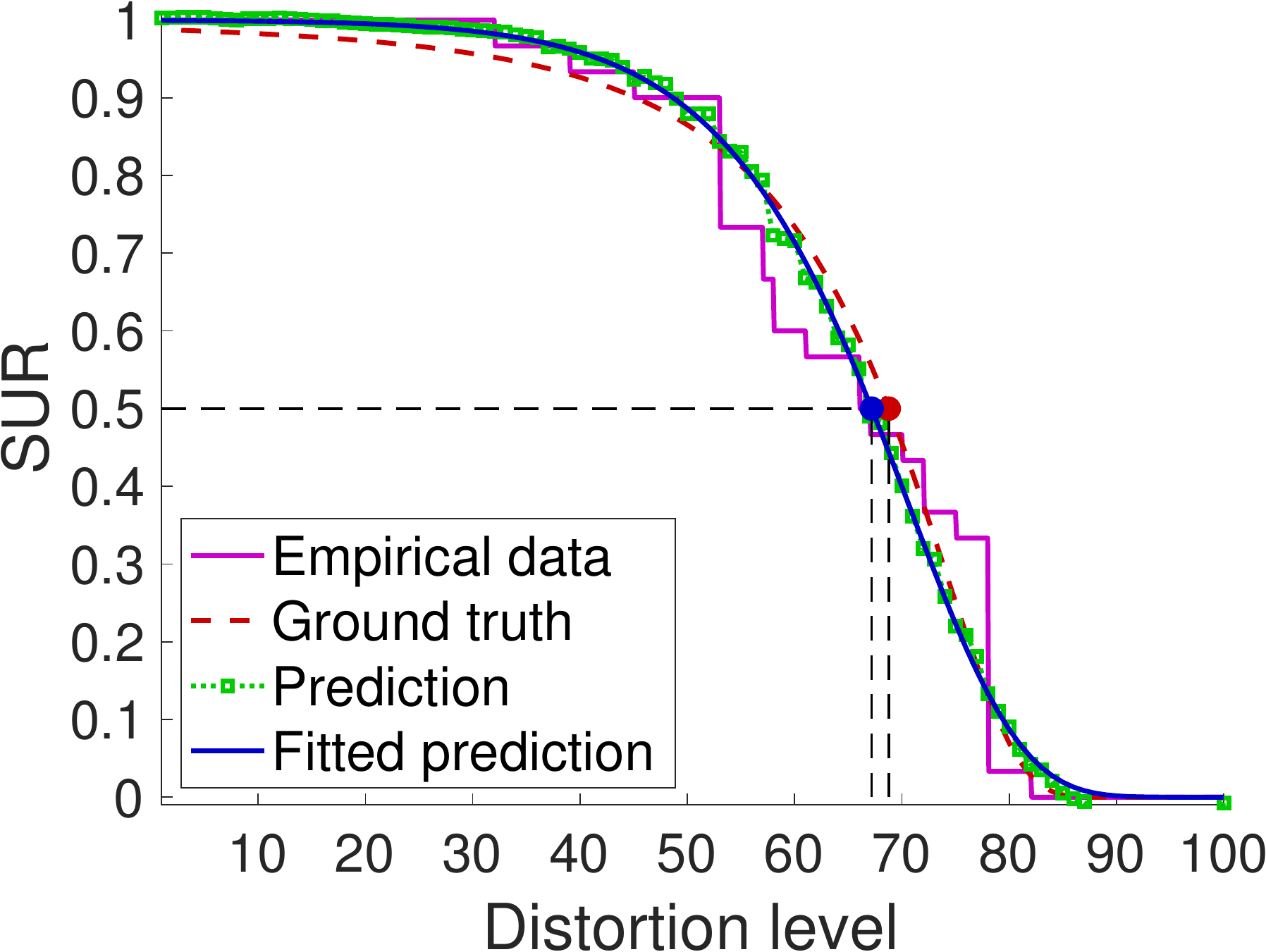}
    \end{minipage}
        \begin{minipage}[c]{0.40\textwidth}
        \includegraphics[width=1\linewidth]{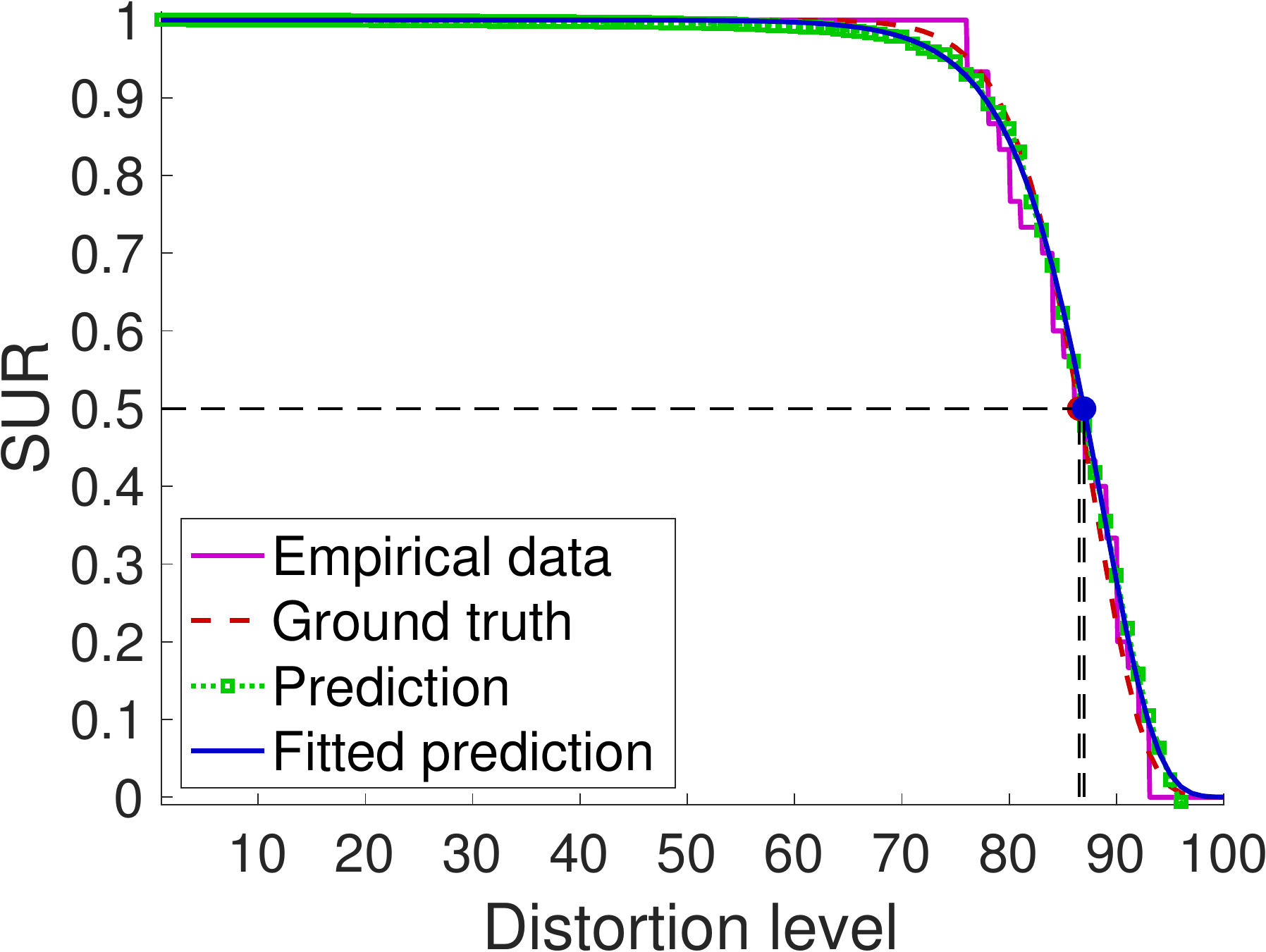}
    \end{minipage}
    \begin{minipage}[c]{0.4\textwidth}
        \includegraphics[width=1\linewidth]{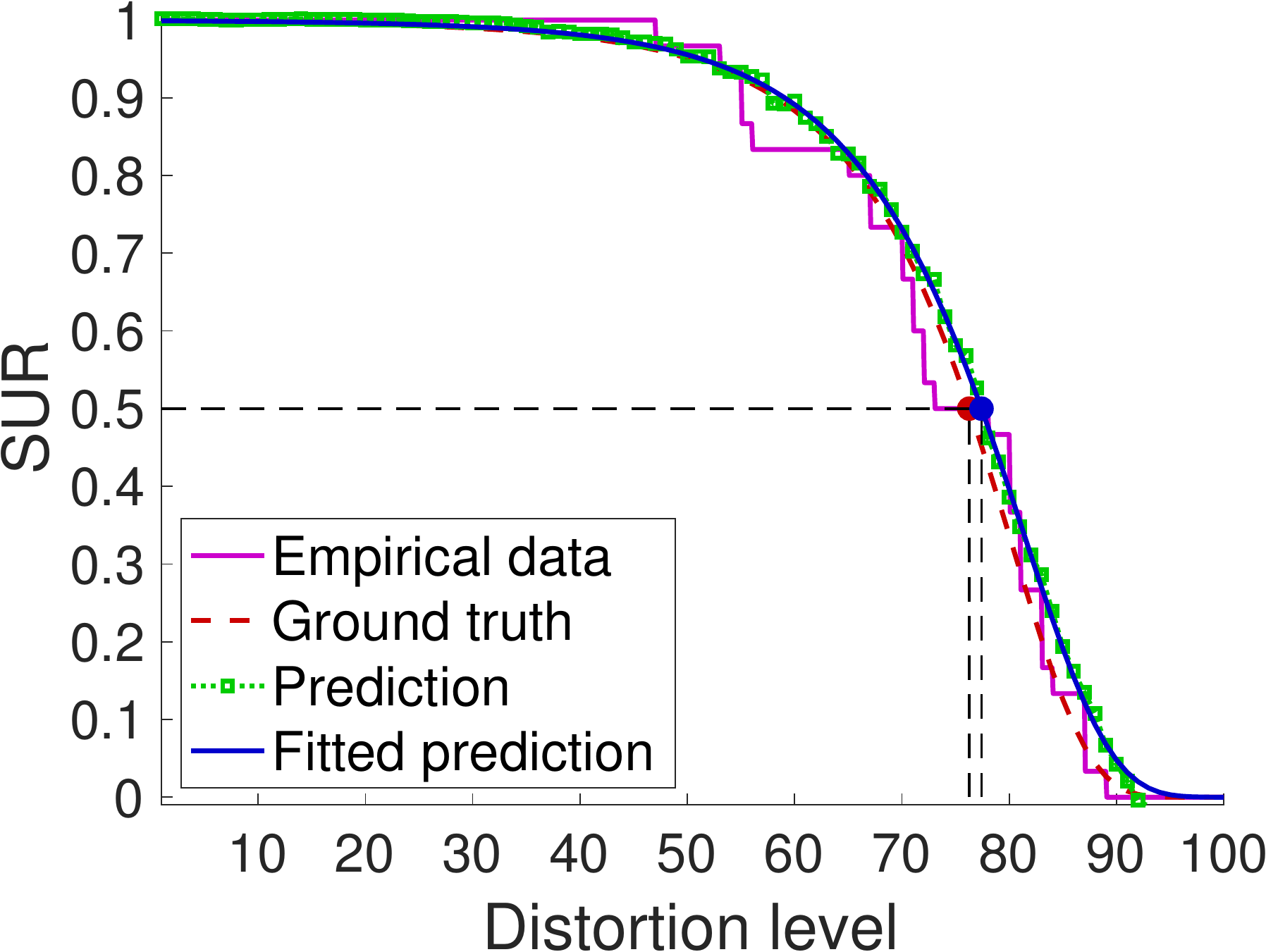}
    \end{minipage}
        \begin{minipage}[c]{0.40\textwidth}
        \includegraphics[width=1\linewidth]{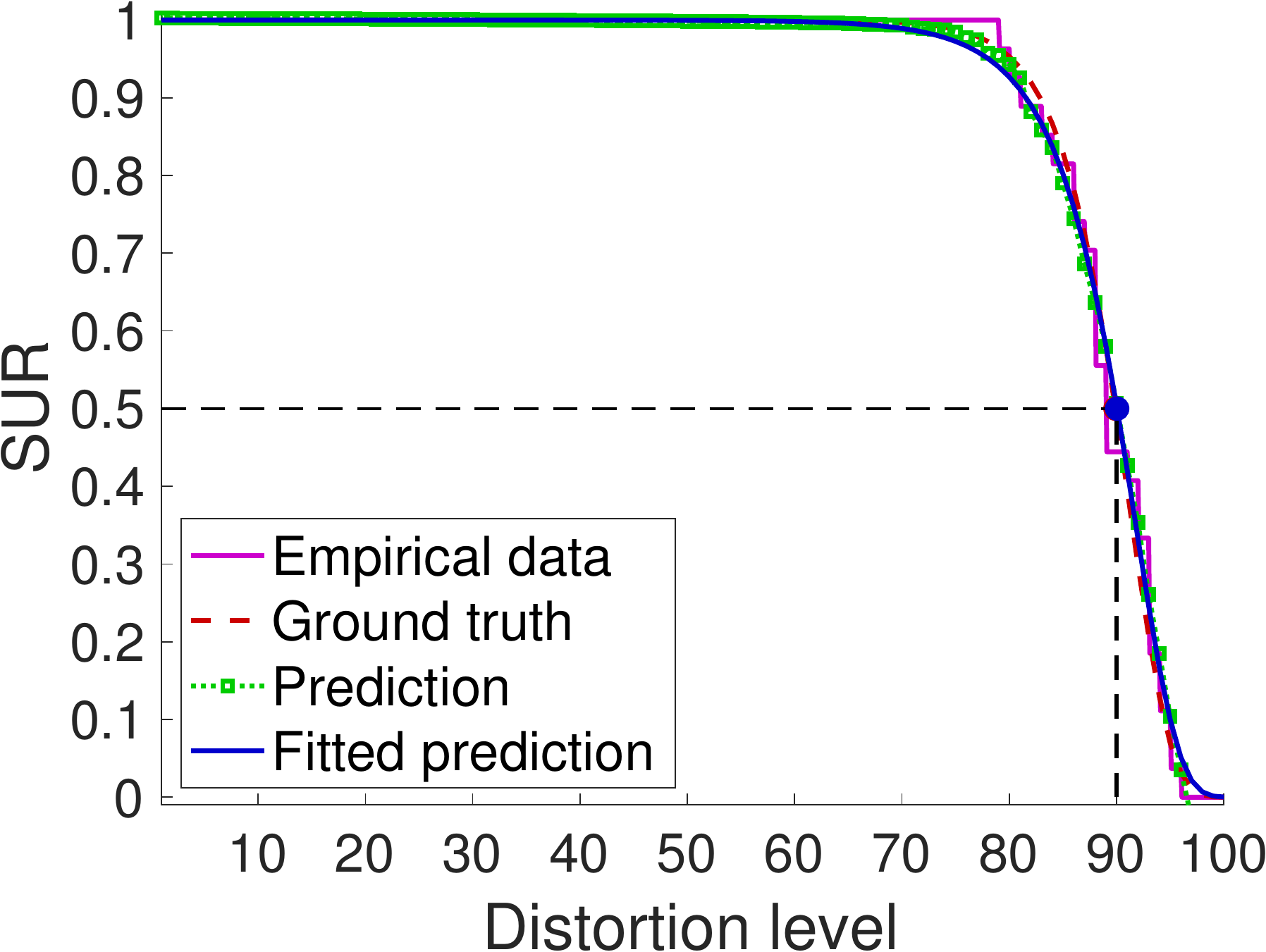}
        \centerline{(a) Best prediction}
    \end{minipage}
    \begin{minipage}[c]{0.40\textwidth}
        \includegraphics[width=1\linewidth]{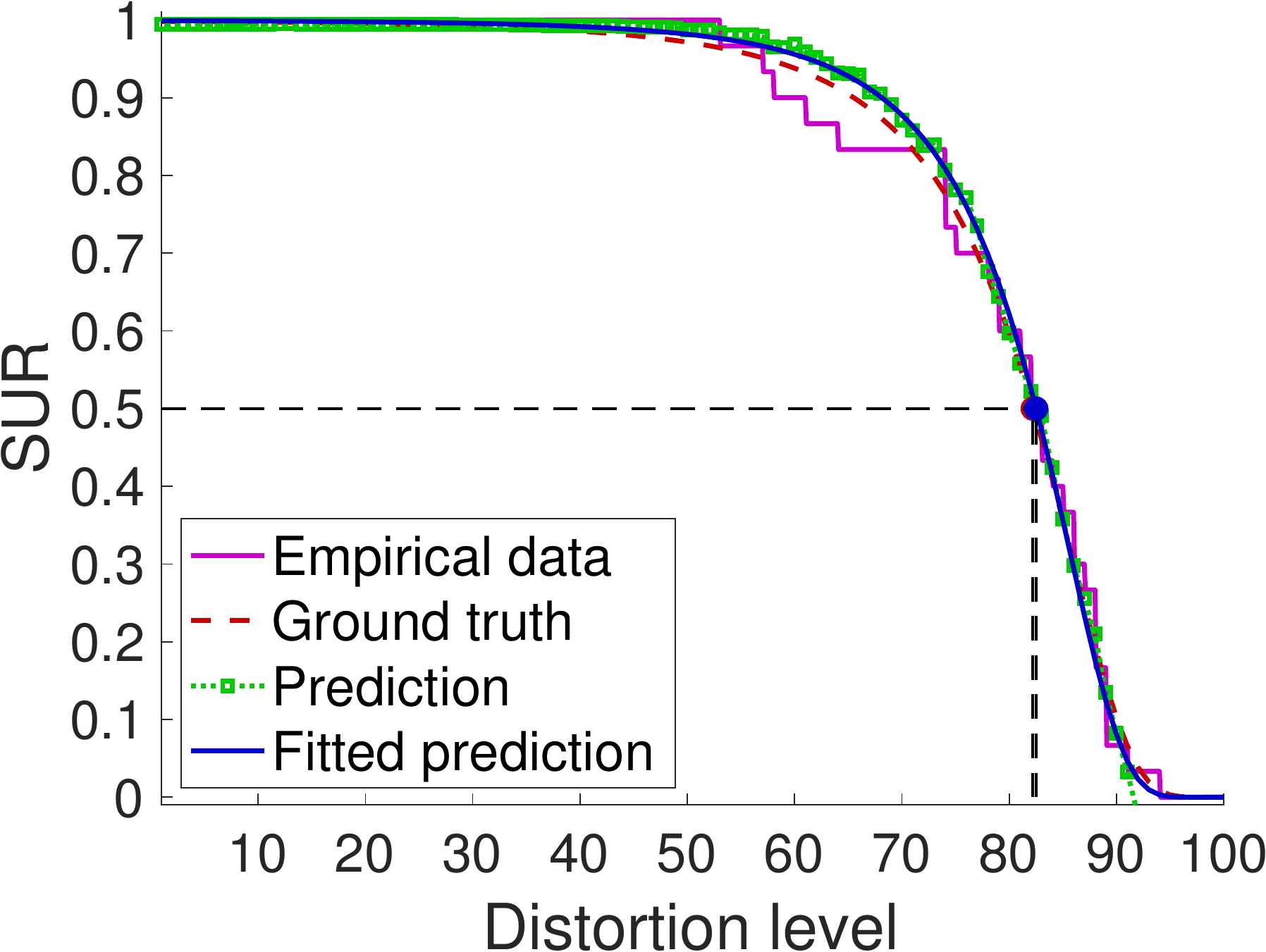}
        \centerline{(b) Second best prediction}
    \end{minipage}
    \caption{Best two prediction results according to overall Bhattacharyya distance for the first three JNDs. The first row shows the source images. The second, third, and fourth rows correspond to the first, second, and third JNDs.} 
\label{fig:besttwopredict}
\end{figure*}

Fig.~\ref{fig:besttwopredict} shows the best two predictions, sorted according to the mean Bhattacharyya distance over the three JNDs. The best prediction result was for image 35, with absolute 50\% JND errors of 0, 0, and 1, Bhattacharyya distances of 0.0073, 0.0073, and 0.0052, and PSNR differences at the 50\% JNDs of 0, 0, and 0.3 dB for the first, second, and third JND, respectively. 

\begin{figure*}[p]
    \centering
    \begin{minipage}[c]{0.40\textwidth}
        \includegraphics[width=1.0\linewidth]{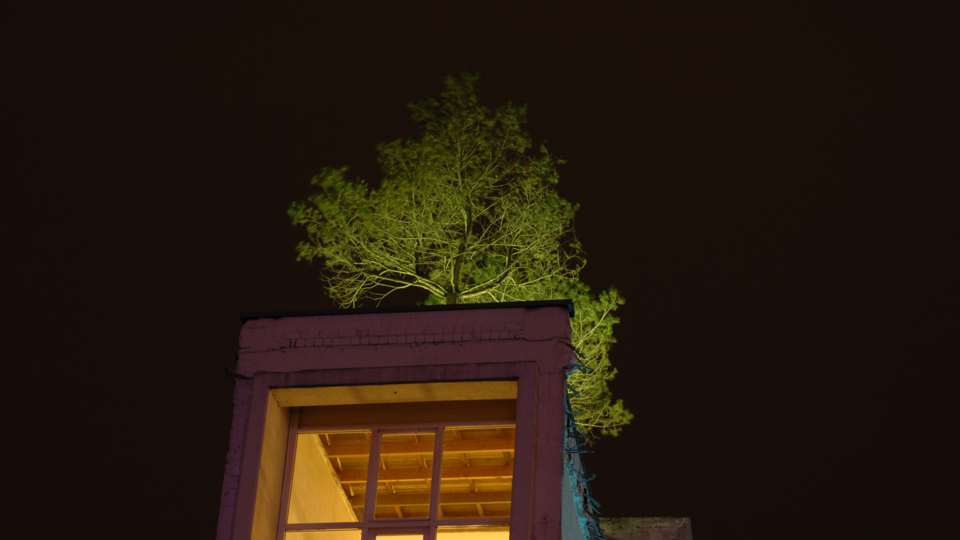}
            \centerline{Image 2}
    \end{minipage}
    \begin{minipage}[c]{0.40\textwidth}
        \includegraphics[width=1.0\linewidth]{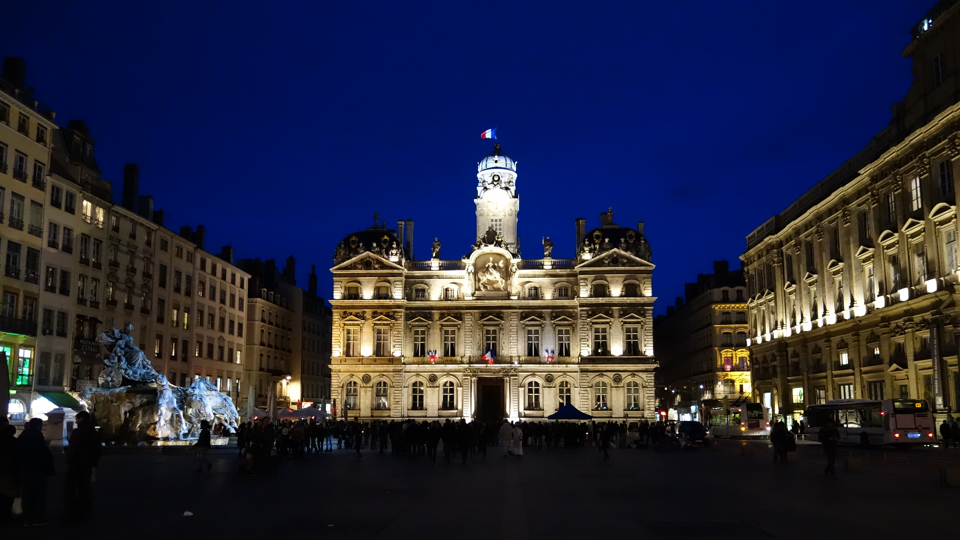}
         \centerline{Image 12}
    \end{minipage}
    \begin{minipage}[c]{0.40\textwidth}
        \includegraphics[width=1\linewidth]{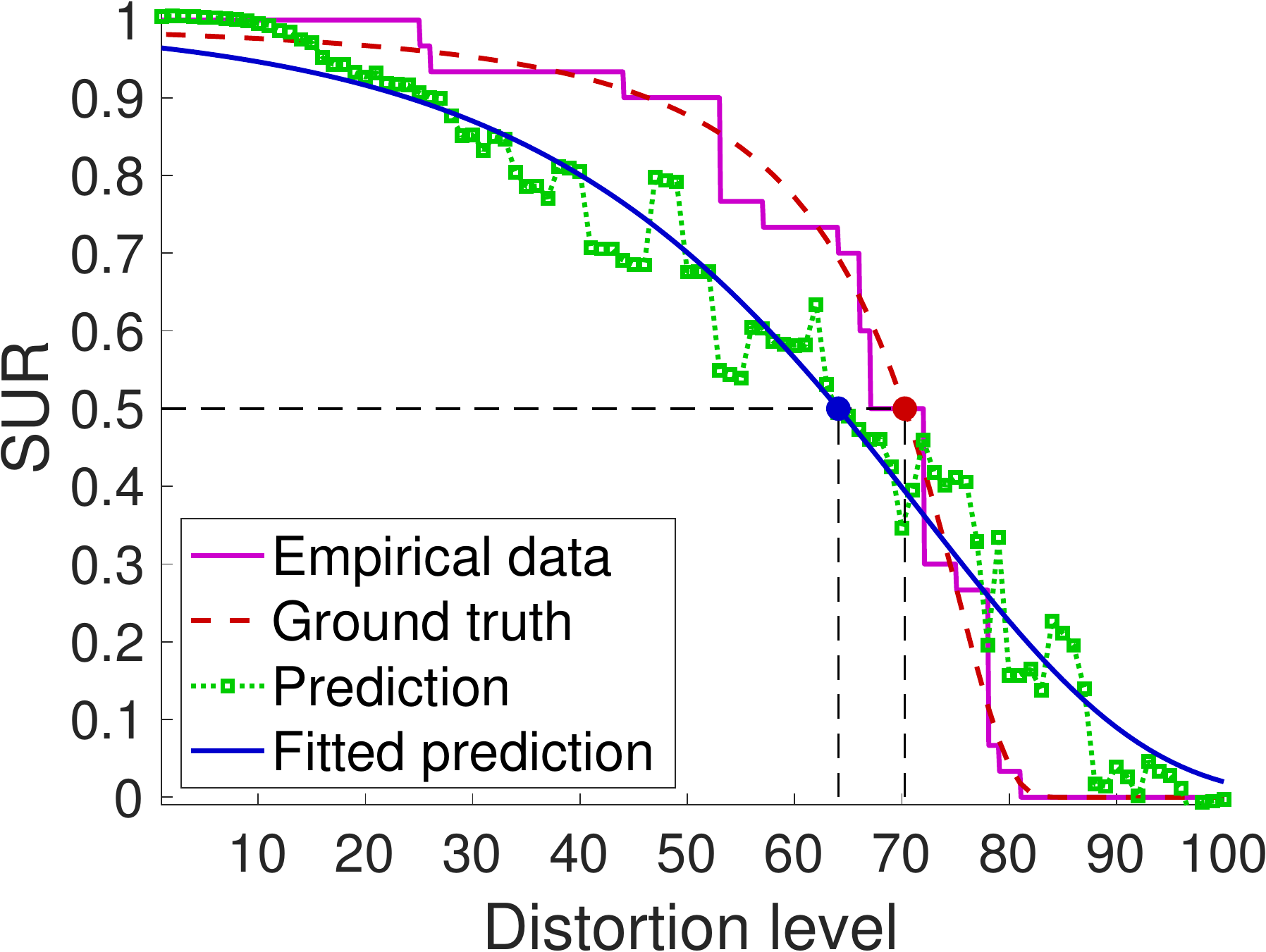}
    \end{minipage}
    \begin{minipage}[c]{0.40\textwidth}
        \centering
        \includegraphics[width=1\linewidth]{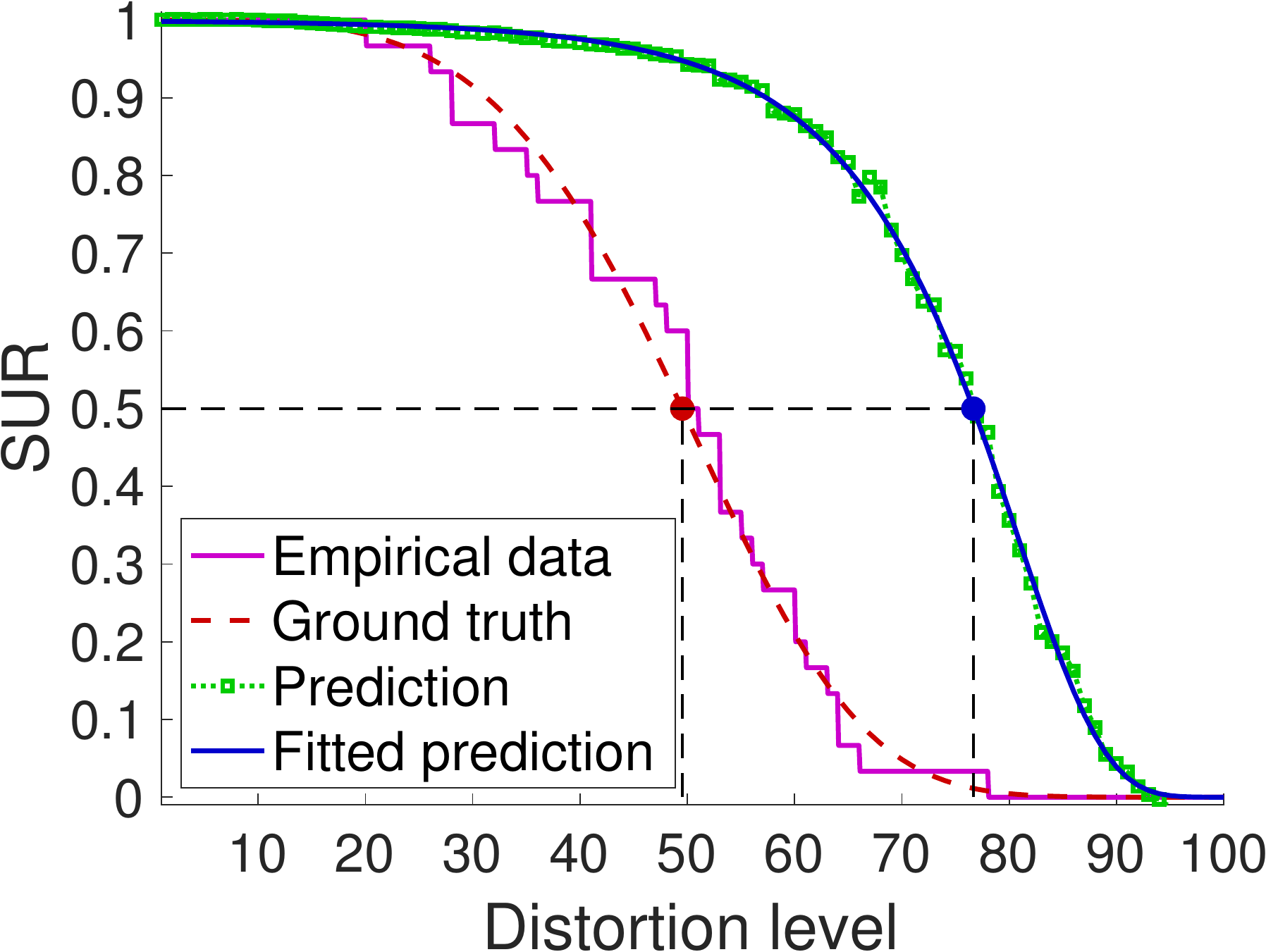}
    \end{minipage}
    \begin{minipage}[c]{0.40\textwidth}
        \includegraphics[width=1\linewidth]{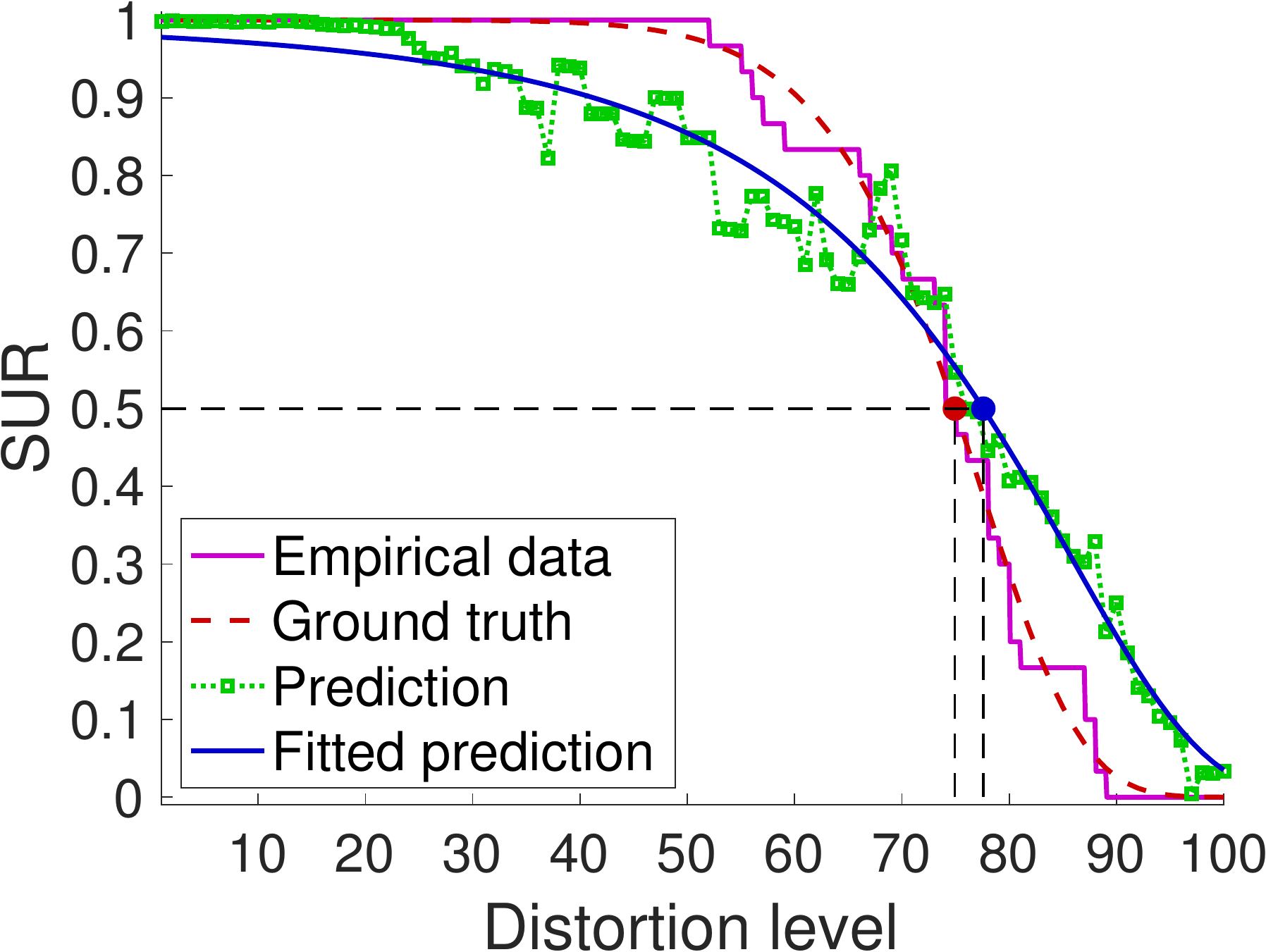}
    \end{minipage}
    \begin{minipage}[c]{0.40\textwidth}
        \includegraphics[width=1\linewidth]{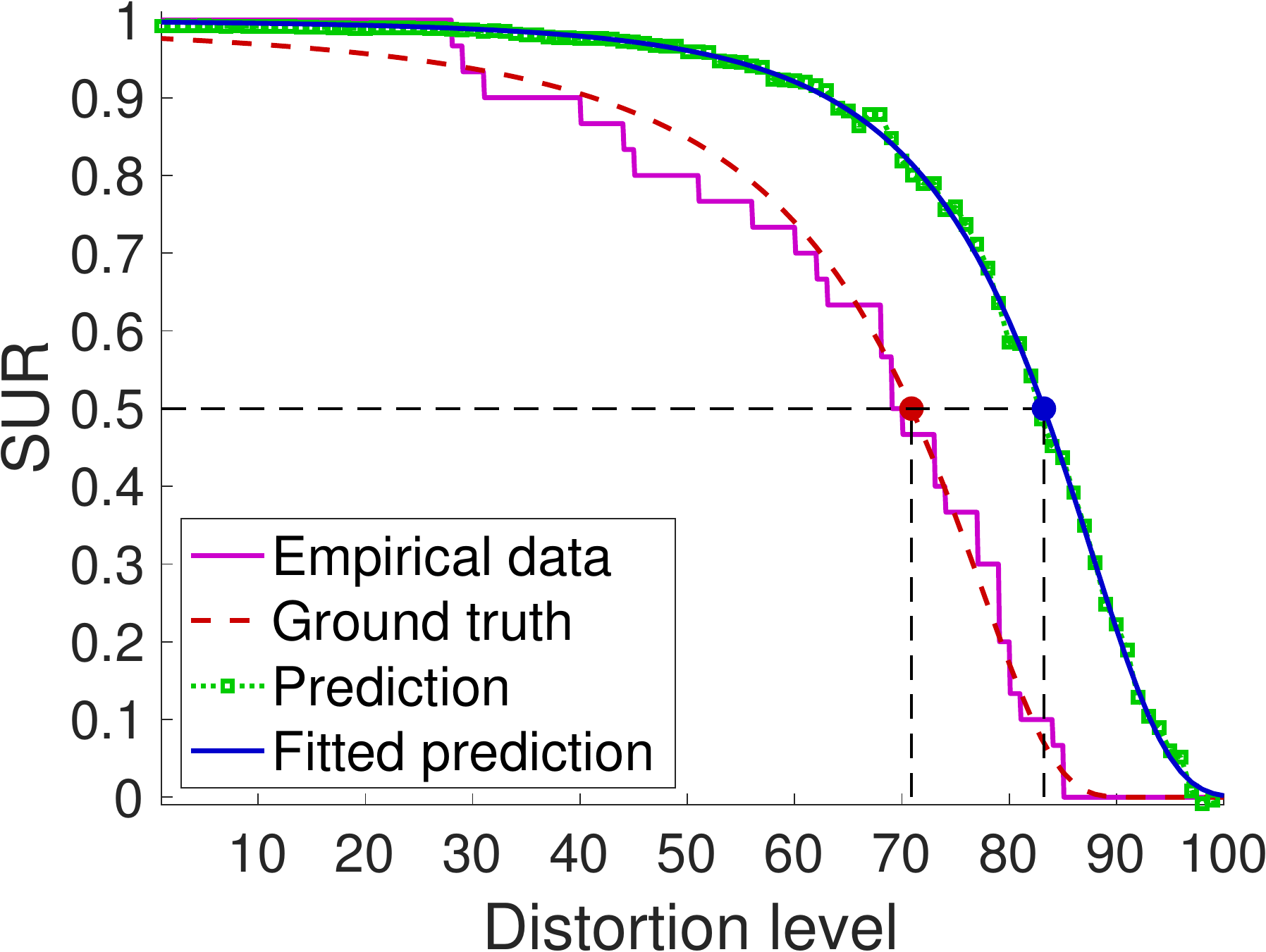}
    \end{minipage}
    \begin{minipage}[c]{0.40\textwidth}
        \includegraphics[width=1\linewidth]{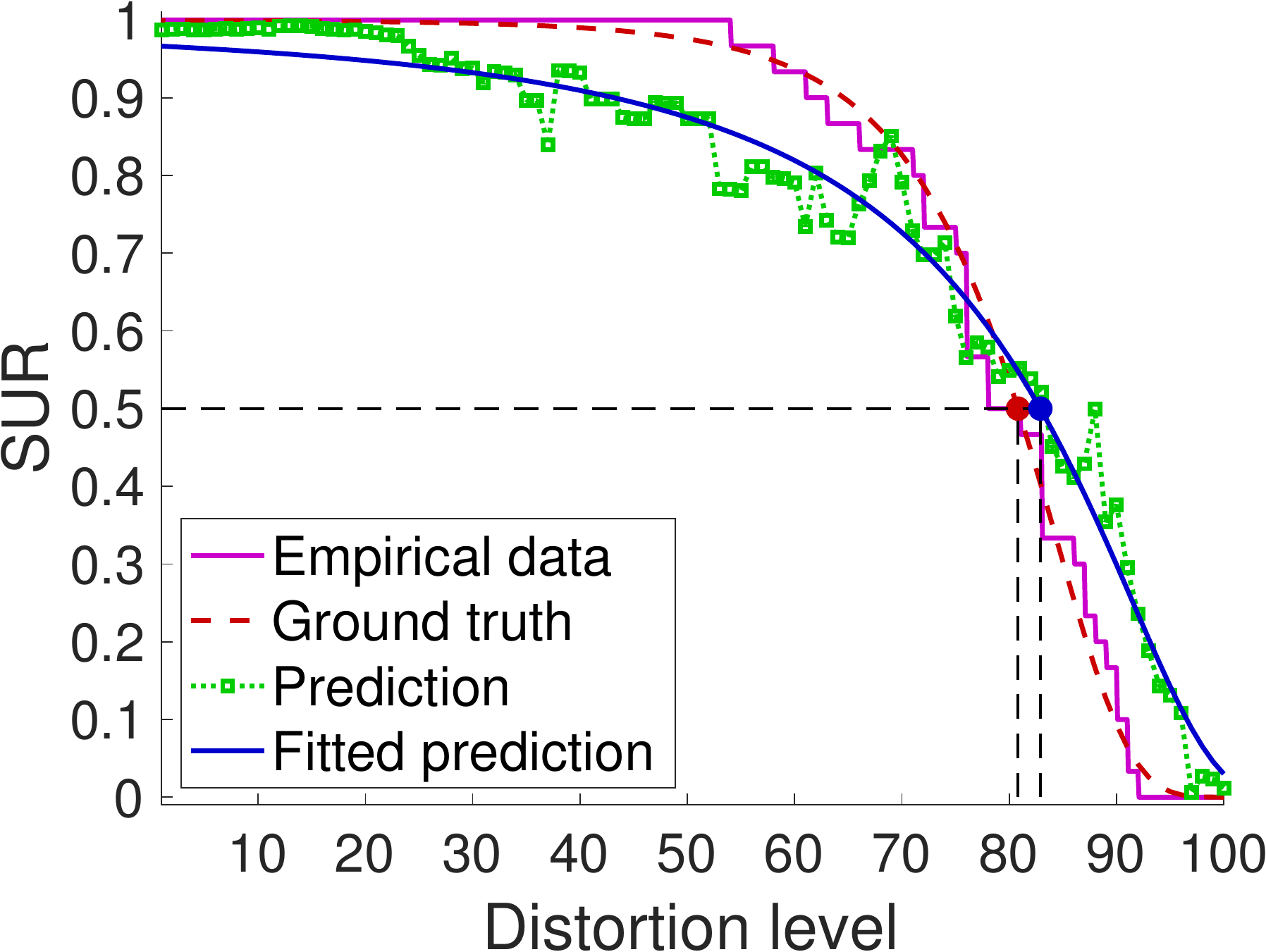}
        \centerline{(a) Second worst prediction}
    \end{minipage}
    \begin{minipage}[c]{0.40\textwidth}
        \includegraphics[width=1\linewidth]{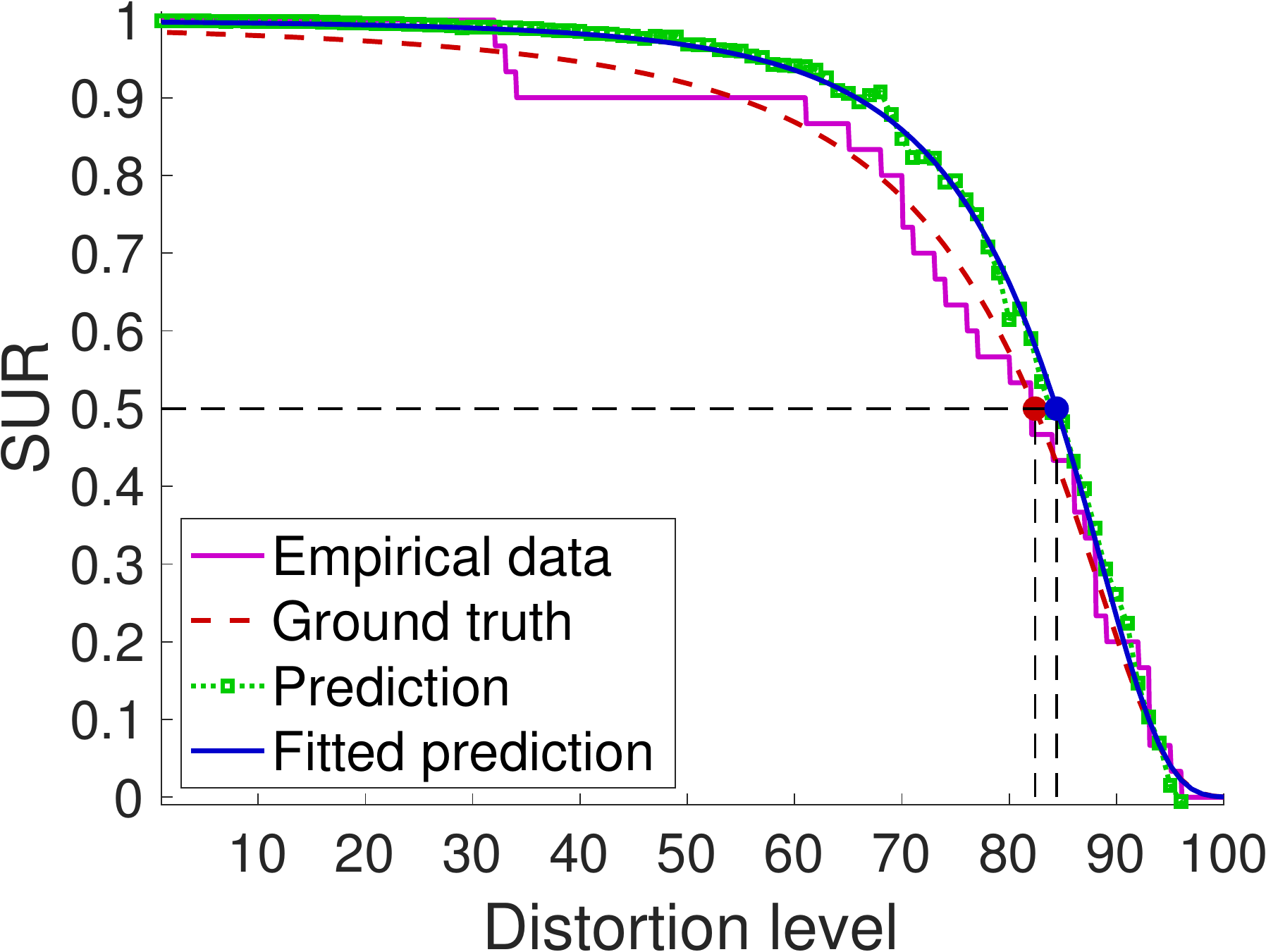}
        \centerline{(b) Worst prediction}
    \end{minipage}
    \caption{Worst two prediction results according to overall Bhattacharyya distance for the first three JNDs. The first row shows the source images. The second, third, and fourth rows correspond to the first, second, and third JNDs.} 
\label{fig:worsetwopredict}
\end{figure*}

The prediction results for a few images were not as good.
For example, Fig~\ref{fig:worsetwopredict} presents the worst two predictions. The  worst prediction was for image 12, which had  absolute 50\% JND errors of 27, 13, and 2, Bhattacharyya distances of 0.4884, 0.2373, and 0.0167, and PSNR differences at the 50\% JND of 2.55, 1.88, and 0.34 for the first, second, and third JND, respectively. This may be because the size and diversity of the training set are too small for the deep learning algorithm. We expect that this problem can be overcome by training on a large-scale JND dataset. 

The overall performance of SUR-FeatNet is displayed in Table~\ref{tb:surnet_perf}. The mean Bhattacharyya distances between the predicted and the ground truth first, second, and third JND distributions were only 0.0810, 0.0702, and 0.0522, respectively. 

\begin{algorithm}
\caption{Baseline method for $p$\% JND prediction}\label{baseline}
\begin{algorithmic}[1]
  \State \textbf{image} $I$                   \Comment{Original test image}
  \State \textbf{float} PSNR$[1...N]$      \Comment{PSNR at $p$\% JNDs in training set}
  \State $T \gets \frac{1}{N}\sum_{n=1}^N \text{PSNR}[n]$     \Comment{PSNR threshold at $p$\% JND}
\Function{JND}{$I,T$} 
   \State $D \gets 0$     \Comment{Initialize distortion level}
   \Repeat
   \State $D \gets D + 1$     \Comment{Increment distortion level}
   \State QF $\gets 101 - D$  \Comment{JPEG quality factor QF}
   \State $\hat{I} \gets $JPEG$^{-1}$(JPEG($I$,QF)) \Comment{Encode / decode}
\Until{PSNR($\hat{I},I) \le T$}
   \State \textbf{return} $D$\Comment{The $p$\% JND is at level $D$.}
\EndFunction
\end{algorithmic}
\end{algorithm}

\begin{figure*}[!t]
\centering
\begin{minipage}{0.32\textwidth}
\includegraphics[width=1.0\linewidth]{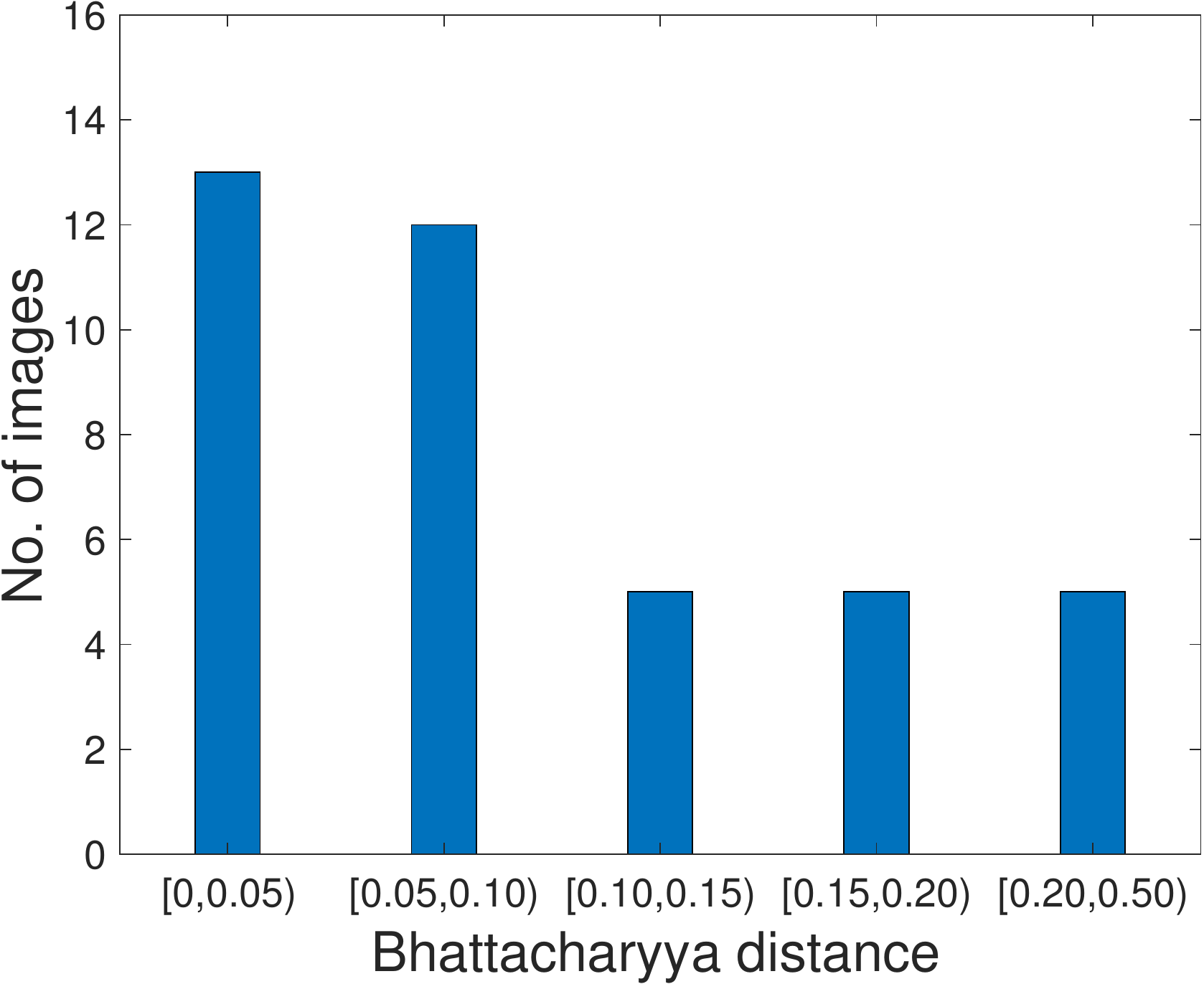}
\centerline{(a)}
\end{minipage} 
\begin{minipage}{0.32\textwidth}
\includegraphics[width=1.0\linewidth]{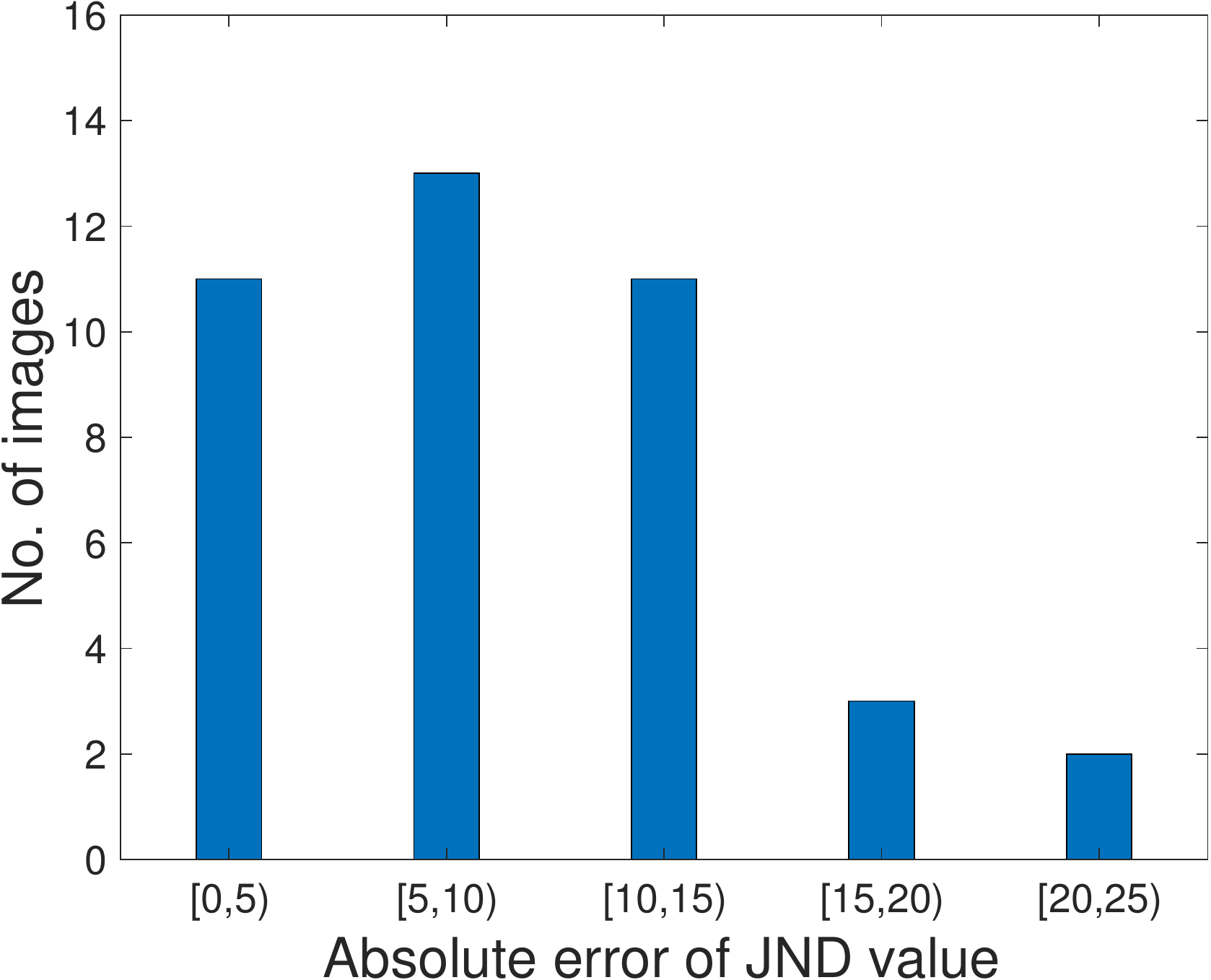}
\centerline{(b)}
\end{minipage}
\begin{minipage}{0.32\textwidth}
\includegraphics[width=1.0\textwidth]{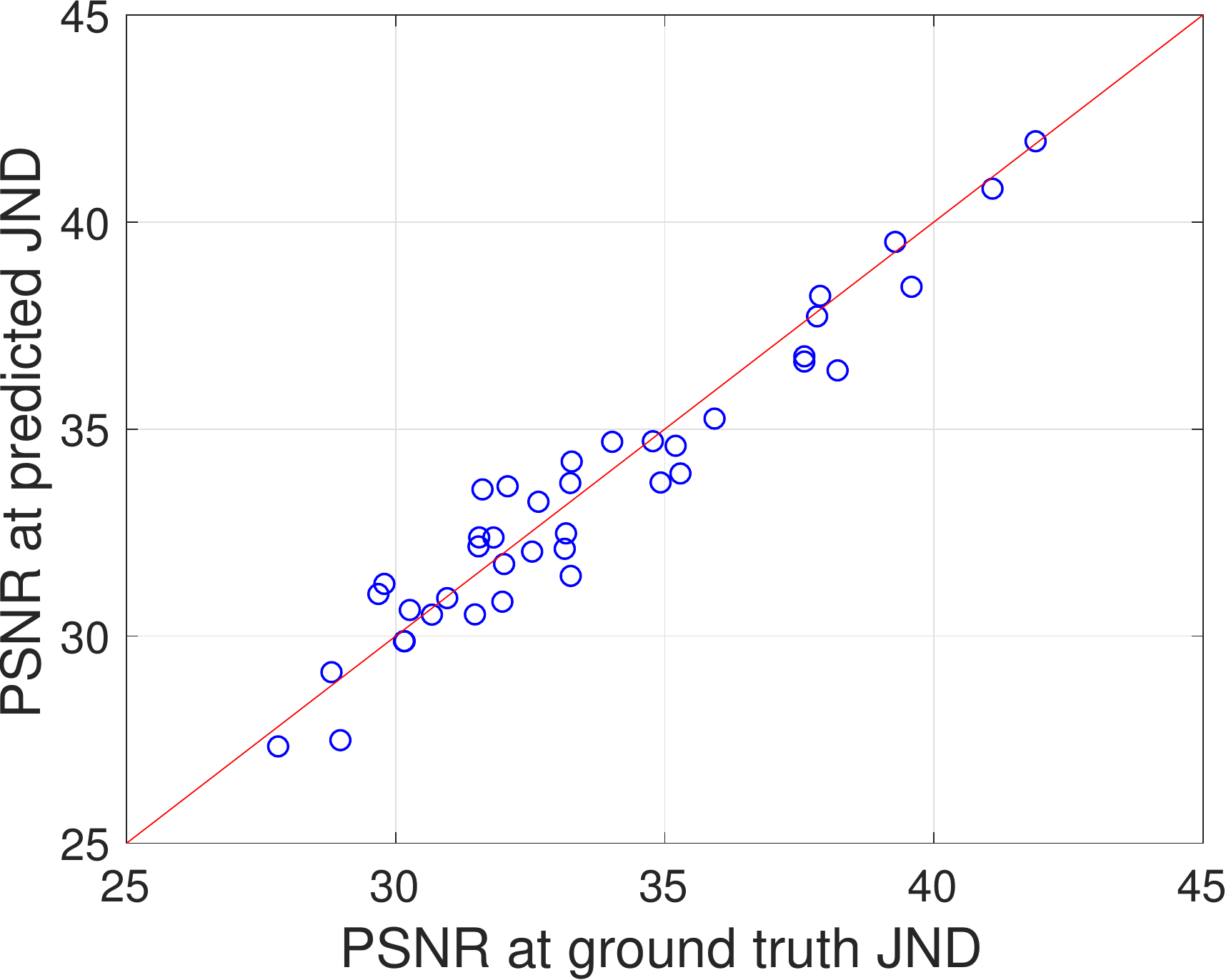}
\centerline{(c)}
\end{minipage}
\caption{\hilite{Statistics of experimental results for the JND-Pano dataset. (a) Histogram of Bhattacharyya distance between predicted JND distribution and ground truth JND distribution. (b) Histogram of the absolute error between predicted JND (50\% JND) and ground truth JND (50\% JND). (c) PSNR comparison between the ground truth JNDs and the predicted JNDs; the PLCC is 0.9651.}
}
\label{fig:pano_stat}
\end{figure*}

\begin{table*}[!htbp]
\centering
\begin{tabular}{l l | c  c  c | c c}
\multicolumn{2}{c}{} & \multicolumn{3}{c}{SUR-FeatNet} & \multicolumn{2}{c}{Baseline}\\ 
& &Bhattacharyya & $\Delta \text{JND}$  & $\Delta \text{PSNR}$  & $\Delta \text{JND}$ & $\Delta \text{PSNR}$ \\
Dataset &JND  &distance &    & (dB) &    & (dB)\\ \hline
\multirow{3}{*}{MCL-JCI} &First &0.0810 & 4.44 & 0.58 &27.97 &2.94 \\
&Second & 0.0702 & 3.34 & 0.69 &31.68 &2.96\\
&Third & 0.0522 & 2.10 & 0.58 &34.14 &2.86\\ \hline
\hilite{JND-Pano} &First & 0.1053 & 8.63 & 0.76 &27.05 &2.93 \\ \hline
\end{tabular}
\caption{\hilite{Comparison between SUR-FeatNet and the baseline method (Algorithm \ref{baseline}) for the two benchmark datasets. $\Delta \text{JND}$ is the MAE of the 50\% JNDs. $\Delta \text{PSNR}$ is the MAE of the PSNR at the 50\% JNDs.}}
\label{tb:surnet_perf}
\end{table*}

\hilite{These performances can be compared with a simple baseline prediction based on the average PSNR at the 50\% JND. 
For that, we use the same data splitting of the $k$-fold cross validation. For each source image, one subset of compressed images was used for testing, and the remaining nine subsets were joined and used together for ``training''.
The 50\% JNDs for the test set were predicted by the distortion levels corresponding to the average PSNRs at the corresponding 50\% JNDs in the training set (see Algorithm 1 for the details).} 

\hilite{Table \ref{tb:surnet_perf} reports the average prediction errors in terms of distortion levels and PSNR. The images in the MCL-JCI dataset corresponding to the 50\% JNDs predicted by the baseline method show an average error in PSNR close to 3~dB while those predicted by SUR-FeatNet are much smaller, ranging from 0.58 to 0.69~dB.}

\subsection{Results and analysis for JND-Pano}

\hilite{The overall performance on the JND-Pano dataset is summarized in Table \ref{tb:surnet_perf}. The average Bhattacharyya distance is 0.1053, the absolute JND error is 8.63, and the PSNR difference at the JND is 0.76 dB. This demonstrates that our SUR-FeatNet also works well for panoramic images and head-mounted displays.} 

\hilite{Nevertheless, the performance for the JND-Pano dataset is not as good as that for MCL-JCI. This is because the JND-Pano dataset is different in character compared to the MCL-JCI dataset: images are panoramic, thus much larger in resolution, and the JND samples are obtained using a different modality, i.e., head-mounted displays rather than screen images. As a result, participants of subjective JND studies for JND-Pano may be more likely to overlook differences between reference and distorted images. This is supported by an analysis of the JND measurements across all images, which yielded an average standard deviation of 14.93 in JND-Pano, compared to only 10.16 in MCL-JCI.}

\hilite{Table~{\ref{Table:pano_first_jnd_result}} presents the detailed results w.r.t.~the first JND for each image sequence in the JND-Pano dataset, and the statistics are shown in Fig~{\ref{fig:pano_stat}}.}

\subsection{Comparison with previous work}

SUR-FeatNet outperformed the state-of-the-art PW-JND model of Liu~\textit{et al.}~\cite{Liujnd2020} for the first and second JND, see Table \ref{tb:stateofart}, except for the mean absolute error of the predicted distortion level of the second JND. There are no results listed in \cite{Liujnd2020} for the third JND. Note, that in \cite{Liujnd2020} the ground truth JNDs are slightly different from those used for SUR-FeatNet, as they had been taken from the model in \cite{MCL-JCI} (Table 2). 
One advantage of our method compared to the work in \cite{Liujnd2020} is that it can predict the distortion level at arbitrary percentiles (e.g., at the 75\% SUR).


SUR-FeatNet also showed a better performance when predicting the 75\% SUR for the first JND in the MCL-JCI dataset, compared to our previous model SUR-Net \cite{fan2019net} (Table \ref{tb:stateofart}).

\begin{table*}[h]
\centering
\begin{tabular}{l  |cc|cc|cc }
\multicolumn{1}{c}{} &
\multicolumn{2}{c}{First JND (50\%JND)} & 
\multicolumn{2}{c}{Second JND (50\%JND)} &
\multicolumn{2}{c}{First JND (75\%SUR)} \\
 & $\Delta \text{JND}$  & $\Delta \text{PSNR}$ & $\Delta \text{JND}$  & $\Delta \text{PSNR}$& $\Delta \text{JND}$  & $\Delta \text{PSNR}$\\
Method &    & (dB) &    & (dB) &    & (dB) \\
\hline
Baseline & 27.97 & 2.94 & 31.68 & 2.96 & -- & --  \\
PW-JND \cite{Liujnd2020} & \phantom{0}8.7\phantom{0} & 0.82 & \phantom{0}\textbf{3.14} & 0.76 & -- & --  \\
SUR-Net \cite{fan2019net} & \phantom{0}5.22 & 0.63 & -- & -- & 6.73 & 0.69 \\ 
\hline
SUR-FeatNet & \phantom{0}\textbf{4.44} & \textbf{0.58} & \phantom{0}3.34 & \textbf{0.69} & \textbf{5.45} & \textbf{0.59} \\ 
\end{tabular}
\caption{Performance comparison with the state-of-the-art PW-JND model \cite{Liujnd2020} and SUR-Net \cite{fan2019net} for the MCL-JCI dataset. $\Delta \text{JND}$ is the MAE with respect to the ground truth. $\Delta \text{PSNR}$ is the MAE of the PSNR.}
\label{tb:stateofart}
\end{table*}

\section{Concluding remarks}

\subsection{Summary}

\hilite{To predict SUR curves, we needed a well-behaved model for the curves themselves. This has led us to search for the best fitting distribution for the empirical JND data. A well-fitting distribution improves the modeling capabilities of any predictive model used subsequently.}

We proposed a deep-learning approach to predict SUR curves for compressed images. In a first stage, pairs of images, a reference and a distorted, are fed into a Siamese CNN to predict an objective quality score. In a second stage, extracted MLSP features are fed into a shallow regression network to predict the SUR value of a given image pair.

For a target percentage of satisfied users, the predicted SUR curve can be used to determine the JPEG quality factor QF that provides a compressed image, which is indistinguishable from the original for these users, thereby providing bitrate savings without the need for subjective visual quality assessment.

\subsection{Limitations and future work}

\hilite{The performance of our model is limited by the small amount of annotated data available. A large scale JND dataset with a large number of diverse-content reference images would significantly improve the performance of our model, as well as other potential models.}

\hilite{We assumed that the image compression scheme is lossy and produces monotonically increasing distortions as a function of an encoding parameter. 
For input images that are noisy, compression at high bitrates may smooth the images, leading to a higher perceptual image quality. Consequently, the psychometric function associated to the distortion will not fit well with the applied model.
}

\hilite{Our model makes independent predictions for each pair of reference and JPEG compression level. The integration of these predictions is implemented as an additional step. A model that is aware of the relations between the predictions at training time could have a better performance. One option would be to directly predict the parameters of the analytical distribution given only the source image, or the source image and all its distorted versions, at the same time. Such an approach may need more training data.}

The proposed method can be easily generalized to predict the SUR curves for images compressed with other encoding methods, or different distortion types. 

We provided results for 50\% JND and 75\% SUR. Results for other percentages can be obtained in a similar way. 

\section*{Conflict of interest}

The authors declare that they have no conflict of interest.

\bibliographystyle{spmpsci}      

\bibliography{reference}


\begin{table*}[p]
\centering
\caption{GEV distribution model of the \textit{first} JND for the 50 image sequences of the MCL-JCI dataset. Shown are the location $\mu$, scale $\sigma$ and shape $\xi$, for both ground truth and SUR-FeatNet, together with the 50\% JND values and PSNR at the 50\% JND value. The Bhattacharyya distance measures the divergence between the predicted and ground truth distributions, $\Delta \text{JND} = |\widehat{\text{JND}} - \text{JND}|$, and  $\Delta \text{PSNR} = |\widehat{\text{PSNR}} - \text{PSNR}|$. 
}
\resizebox{0.99\textwidth}{!}{
\begin{tabular}{c|rrrrr|rrrrr|c|r|r}
\hline
Image & \multicolumn{5}{c|}{Ground truth}   & \multicolumn{5}{c|}{SUR-FeatNet}   & Bhattacharyya & \multicolumn{1}{c|}{$\Delta \text{JND}$} & \multicolumn{1}{c}{$\Delta \text{PSNR}$}\\
$k$ & \multicolumn{1}{c}{$\mu$} & \multicolumn{1}{c}{$\sigma$} & \multicolumn{1}{c}{$\xi$} & \multicolumn{1}{c}{\text{JND}} & \multicolumn{1}{c|}{\text{PSNR}} & \multicolumn{1}{c}{$\hat{\mu}$} & \multicolumn{1}{c}{$\hat{\sigma}$} & \multicolumn{1}{c}{$\hat{\xi}$} & \multicolumn{1}{c}{$\widehat{\text{JND}}$} & \multicolumn{1}{c|}{$\widehat{\text{PSNR}}$} & distance & & (dB)\\ \hline
1 &22.61 &6.36 &-0.15 &77 &31.94 &18.62 &7.47 &0.25 &80 &31.42 &0.0781 &3 &0.52 \\
2 &27.82 &7.36 &0.40 &71 &39.84 &29.25 &20.88 &0.01 &65 &40.75 &0.1964 &6 &0.92 \\
3 &22.53 &8.50 &0.28 &76 &31.47 &23.73 &8.83 &0.16 &74 &31.70 &0.0105 &2 &0.23 \\
4 &21.30 &5.36 &0.18 &78 &28.77 &20.33 &8.15 &0.11 &78 &28.77 &0.0514 &0 &0.00 \\
5 &24.29 &3.94 &0.55 &76 &31.58 &24.60 &9.22 &0.01 &74 &31.86 &0.1469 &2 &0.28 \\ \hline
6 &22.30 &4.29 &0.73 &77 &32.96 &20.35 &5.93 &0.10 &79 &32.66 &0.1476 &2 &0.30 \\
7 &31.98 &12.22 &0.14 &65 &29.92 &25.07 &9.83 &-0.07 &73 &29.02 &0.0735 &8 &0.90 \\
8 &23.79 &3.80 &0.29 &76 &28.31 &26.35 &9.82 &-0.05 &72 &28.79 &0.1341 &4 &0.47 \\
9 &17.53 &3.97 &0.13 &82 &27.97 &20.61 &5.37 &-0.02 &79 &28.43 &0.0444 &3 &0.46 \\
10 &21.04 &4.08 &0.39 &79 &36.45 &23.08 &10.20 &0.14 &75 &37.05 &0.1264 &4 &0.59 \\ \hline
11 &31.15 &9.35 &-0.19 &67 &34.34 &22.24 &5.77 &-0.07 &77 &33.31 &0.1739 &10 &1.03 \\
12 &46.97 &12.76 &-0.22 &50 &34.11 &20.98 &8.99 &0.10 &77 &31.56 &0.4884 &27 &2.55 \\
13 &21.25 &4.97 &0.06 &78 &35.38 &20.76 &7.18 &0.18 &78 &35.38 &0.0384 &0 &0.00 \\
14 &20.79 &7.26 &0.01 &78 &32.90 &19.90 &9.48 &-0.12 &78 &32.90 &0.0216 &0 &0.00 \\
15 &20.53 &7.65 &0.20 &78 &26.45 &15.22 &5.46 &0.11 &84 &25.78 &0.0830 &6 &0.67 \\ \hline
16 &19.66 &8.33 &0.11 &79 &30.35 &21.99 &8.00 &0.07 &77 &30.61 &0.0133 &2 &0.26 \\
17 &15.41 &5.30 &0.23 &84 &28.64 &16.38 &6.00 &-0.01 &83 &28.80 &0.0140 &1 &0.16 \\
18 &19.42 &6.87 &-0.29 &80 &33.34 &18.34 &6.40 &-0.03 &81 &33.16 &0.0215 &1 &0.19 \\
19 &22.18 &8.23 &-0.17 &76 &29.59 &32.46 &9.58 &-0.05 &66 &30.72 &0.1861 &10 &1.13 \\
20 &39.41 &11.21 &-0.70 &58 &32.62 &29.75 &10.56 &-0.17 &68 &31.69 &0.1118 &10 &0.94 \\ \hline
21 &33.01 &11.76 &-0.40 &64 &29.49 &29.53 &9.58 &0.06 &68 &29.07 &0.0552 &4 &0.41 \\
22 &20.45 &8.65 &0.22 &78 &28.63 &19.17 &8.38 &0.12 &79 &28.52 &0.0100 &1 &0.12 \\
23 &20.13 &3.69 &-0.01 &80 &26.56 &22.23 &7.23 &-0.00 &77 &26.93 &0.0934 &3 &0.38 \\
24 &21.03 &6.06 &-0.06 &78 &32.66 &18.98 &6.47 &0.08 &80 &32.38 &0.0181 &2 &0.28 \\
25 &21.57 &8.27 &-0.15 &77 &28.92 &15.86 &5.82 &-0.06 &84 &27.80 &0.0826 &7 &1.11 \\ \hline
26 &39.44 &11.83 &-1.38 &59 &33.98 &30.39 &13.27 &-0.13 &66 &33.34 &0.3194 &7 &0.64 \\
27 &15.55 &6.61 &-0.08 &84 &29.10 &16.99 &6.77 &0.04 &82 &29.48 &0.0128 &2 &0.38 \\
28 &23.77 &6.66 &0.36 &75 &39.79 &25.07 &9.39 &-0.13 &73 &40.15 &0.0597 &2 &0.35 \\
29 &23.07 &5.65 &-0.01 &76 &36.16 &16.32 &8.56 &-0.18 &82 &35.30 &0.1595 &6 &0.86 \\
30 &18.90 &6.99 &0.01 &80 &35.34 &16.91 &5.57 &0.20 &82 &34.90 &0.0163 &2 &0.44 \\ \hline
31 &21.29 &7.97 &0.03 &77 &33.47 &18.85 &7.39 &0.23 &80 &32.96 &0.0151 &3 &0.51 \\
32 &22.41 &6.81 &0.05 &77 &30.98 &19.25 &5.76 &0.12 &80 &30.49 &0.0254 &3 &0.49 \\
33 &18.43 &4.80 &0.12 &81 &32.16 &23.00 &7.47 &-0.02 &76 &32.98 &0.0647 &5 &0.83 \\
34 &26.51 &6.51 &0.18 &73 &31.30 &26.33 &7.94 &-0.09 &72 &31.40 &0.0251 &1 &0.10 \\
35 &20.20 &6.86 &0.09 &79 &30.74 &19.55 &7.70 &0.16 &79 &30.74 &0.0073 &0 &0.00 \\ \hline
36 &20.33 &6.70 &-0.18 &79 &29.73 &21.30 &7.25 &0.03 &78 &29.87 &0.0232 &1 &0.13 \\
37 &43.62 &19.15 &-0.41 &51 &29.80 &25.35 &8.44 &0.13 &73 &27.47 &0.2351 &22 &2.33 \\
38 &18.27 &6.11 &0.15 &81 &29.26 &18.83 &7.68 &-0.21 &80 &29.43 &0.0356 &1 &0.17 \\
39 &21.30 &8.13 &0.13 &77 &33.73 &23.85 &8.42 &0.06 &75 &34.04 &0.0117 &2 &0.31 \\
40 &28.79 &8.96 &0.25 &69 &38.14 &30.12 &10.12 &-0.02 &68 &38.27 &0.0187 &1 &0.13 \\ \hline
41 &16.72 &6.21 &0.19 &82 &27.21 &19.45 &7.23 &0.10 &79 &27.64 &0.0169 &3 &0.43 \\
42 &18.13 &5.68 &0.15 &81 &29.97 &22.67 &7.66 &0.06 &76 &30.71 &0.0473 &5 &0.74 \\
43 &26.92 &8.29 &0.19 &71 &35.63 &21.44 &7.02 &-0.21 &78 &34.70 &0.1413 &7 &0.93 \\
44 &15.25 &3.83 &0.17 &85 &28.85 &16.60 &5.07 &0.07 &83 &29.25 &0.0154 &2 &0.40 \\
45 &41.88 &15.58 &-0.28 &54 &44.46 &35.50 &11.38 &-0.17 &62 &43.62 &0.0448 &8 &0.84 \\ \hline
46 &12.98 &5.77 &0.10 &86 &31.00 &19.46 &6.93 &0.12 &79 &32.46 &0.1235 &7 &1.46 \\
47 &24.49 &10.31 &0.13 &73 &35.29 &23.99 &8.51 &0.10 &74 &35.14 &0.0105 &1 &0.15 \\
48 &16.81 &6.34 &0.23 &82 &33.59 &23.58 &7.46 &0.02 &75 &34.81 &0.0995 &7 &1.22 \\
49 &19.10 &6.59 &0.18 &80 &36.39 &17.36 &6.97 &0.45 &81 &36.25 &0.0245 &1 &0.14 \\
50 &15.71 &4.53 &0.15 &84 &33.15 &20.47 &6.52 &-0.17 &79 &34.24 &0.0791 &5 &1.10 \\ \hline
\multicolumn{11}{r|}{\bf Avg.}& {\bf0.0810 } & {\bf 4.44} & {\bf 0.58} \\ \hline
\end{tabular}
}
\label{Table:first_jnd_result}
\end{table*}

\begin{table*}[!htbp]
\centering
\caption{GEV distribution model of the \textit{second} JND for the 50 image sequences of the MCL-JCI dataset. Shown are the location $\mu$, scale $\sigma$ and shape $\xi$, for both ground truth and SUR-FeatNet, together with the 50\% JND values and PSNR at the 50\% JND value. The Bhattacharyya distance measures the divergence between the predicted and ground truth distributions, $\Delta \text{JND} = |\widehat{\text{JND}} - \text{JND}|$, and  $\Delta \text{PSNR} = |\widehat{\text{PSNR}} - \text{PSNR}|$. 
}
\resizebox{0.99\textwidth}{!}{
\begin{tabular}{c|rrrrr|rrrrr|c|r|r}
\hline
Image & \multicolumn{5}{c|}{Ground truth}   & \multicolumn{5}{c|}{SUR-FeatNet}   & Bhattacharyya & \multicolumn{1}{c|}{$\Delta \text{JND}$} & \multicolumn{1}{c}{$\Delta \text{PSNR}$}\\
$k$ & \multicolumn{1}{c}{$\mu$} & \multicolumn{1}{c}{$\sigma$} & \multicolumn{1}{c}{$\xi$} & \multicolumn{1}{c}{\text{JND}} & \multicolumn{1}{c|}{\text{PSNR}} & \multicolumn{1}{c}{$\hat{\mu}$} & \multicolumn{1}{c}{$\hat{\sigma}$} & \multicolumn{1}{c}{$\hat{\xi}$} & \multicolumn{1}{c}{$\widehat{\text{JND}}$} & \multicolumn{1}{c|}{$\widehat{\text{PSNR}}$} & distance & & (dB)\\ \hline
1 &15.32 &3.84 &-0.22 &85 &30.33 &11.24 &6.90 &0.13 &88 &29.51 &0.1977 &3 &0.82 \\
2 &22.93 &8.70 &-0.09 &75 &39.58 &17.64 &15.26 &0.18 &78 &38.38 &0.1343 &3 &1.20 \\
3 &14.48 &4.36 &0.33 &85 &30.10 &12.87 &6.42 &-0.05 &86 &29.90 &0.0879 &1 &0.20 \\
4 &15.17 &5.36 &0.24 &84 &27.86 &14.31 &5.99 &-0.02 &85 &27.68 &0.0278 &1 &0.18 \\
5 &17.35 &5.52 &0.16 &82 &30.60 &16.52 &7.70 &0.03 &82 &30.60 &0.0355 &0 &0.00 \\ \hline
6 &16.24 &5.13 &-0.13 &83 &31.97 &14.69 &5.55 &0.03 &85 &31.54 &0.0186 &2 &0.43 \\
7 &21.82 &10.98 &0.06 &76 &28.64 &12.82 &7.99 &-0.14 &86 &27.15 &0.1591 &10 &1.48 \\
8 &18.49 &3.04 &0.27 &82 &27.48 &18.67 &7.90 &0.05 &80 &27.78 &0.1520 &2 &0.30 \\
9 &12.24 &2.51 &0.36 &88 &26.82 &15.33 &5.87 &-0.18 &84 &27.63 &0.1360 &4 &0.81 \\
10 &14.52 &3.93 &0.03 &86 &34.99 &15.23 &6.45 &0.07 &84 &35.48 &0.0490 &2 &0.49 \\ \hline
11 &22.21 &7.18 &-0.10 &77 &33.31 &15.96 &6.26 &-0.15 &83 &32.48 &0.1113 &6 &0.83 \\
12 &26.18 &10.06 &0.33 &71 &32.31 &14.53 &8.57 &0.17 &84 &30.43 &0.2373 &13 &1.88 \\
13 &14.00 &3.14 &0.14 &86 &33.77 &14.87 &6.07 &-0.02 &84 &34.26 &0.0747 &2 &0.49 \\
14 &12.49 &4.93 &-0.01 &87 &31.34 &13.10 &6.83 &-0.04 &86 &31.55 &0.0213 &1 &0.20 \\
15 &11.92 &4.37 &0.23 &88 &25.22 &13.47 &5.12 &0.23 &86 &25.52 &0.0125 &2 &0.30 \\ \hline
16 &12.75 &5.36 &0.14 &87 &29.03 &12.60 &6.58 &-0.33 &87 &29.03 &0.0604 &0 &0.00 \\
17 &11.11 &3.74 &-0.02 &89 &27.72 &10.24 &5.04 &-0.27 &89 &27.72 &0.0390 &0 &0.00 \\
18 &12.59 &5.18 &-0.38 &87 &31.81 &14.48 &6.84 &-0.06 &85 &32.34 &0.1072 &2 &0.52 \\
19 &15.06 &6.79 &-0.13 &84 &28.45 &18.38 &9.16 &-0.16 &80 &29.07 &0.0398 &4 &0.62 \\
20 &25.99 &10.56 &-0.48 &72 &31.23 &20.16 &10.21 &-0.22 &78 &30.49 &0.0497 &6 &0.75 \\ \hline
21 &21.08 &9.85 &-0.28 &77 &27.98 &19.96 &8.09 &-0.04 &79 &27.71 &0.0188 &2 &0.27 \\
22 &11.29 &4.04 &0.32 &89 &26.90 &13.05 &6.57 &0.22 &86 &27.49 &0.0357 &3 &0.59 \\
23 &14.87 &4.21 &-0.05 &85 &25.83 &16.93 &6.64 &0.03 &82 &26.28 &0.0557 &3 &0.45 \\
24 &15.37 &5.89 &-0.15 &84 &31.72 &13.07 &7.35 &-0.10 &86 &31.28 &0.0306 &2 &0.44 \\
25 &12.72 &4.69 &0.02 &87 &27.21 &15.73 &7.63 &0.26 &83 &28.00 &0.0692 &4 &0.79 \\ \hline
26 &23.66 &10.61 &-0.32 &74 &32.41 &20.34 &10.21 &-0.16 &78 &31.87 &0.0196 &4 &0.54 \\
27 &11.13 &3.55 &-0.04 &89 &27.89 &10.52 &4.03 &-0.16 &90 &27.60 &0.0123 &1 &0.29 \\
28 &16.23 &6.14 &0.14 &83 &38.18 &16.91 &7.81 &0.02 &82 &38.38 &0.0128 &1 &0.20 \\
29 &17.29 &4.32 &-0.17 &83 &35.12 &12.46 &6.89 &-0.14 &87 &34.18 &0.1357 &4 &0.95 \\
30 &12.21 &5.08 &-0.02 &87 &33.53 &13.26 &5.42 &0.01 &86 &33.85 &0.0058 &1 &0.32 \\ \hline
31 &12.90 &4.55 &0.20 &87 &31.40 &9.51 &7.10 &-0.36 &90 &30.42 &0.1596 &3 &0.98 \\
32 &13.66 &4.22 &-0.03 &86 &29.27 &10.54 &3.84 &-0.04 &90 &28.13 &0.0695 &4 &1.15 \\
33 &11.76 &3.97 &0.13 &88 &30.57 &14.93 &7.69 &-0.03 &84 &31.57 &0.0784 &4 &0.99 \\
34 &16.96 &5.76 &0.17 &82 &30.07 &14.30 &5.35 &-0.09 &85 &29.55 &0.0578 &3 &0.52 \\
35 &12.87 &4.47 &-0.08 &87 &29.16 &12.22 &4.94 &-0.00 &87 &29.16 &0.0073 &0 &0.00 \\ \hline
36 &14.66 &5.98 &-0.06 &85 &28.77 &13.52 &6.17 &0.06 &86 &28.57 &0.0080 &1 &0.20 \\
37 &28.27 &15.47 &-0.10 &68 &28.06 &19.36 &9.31 &0.34 &79 &26.69 &0.0813 &11 &1.38 \\
38 &10.83 &3.94 &0.14 &89 &27.64 &14.54 &5.93 &0.05 &85 &28.55 &0.0620 &4 &0.91 \\
39 &11.32 &5.73 &0.38 &88 &31.43 &16.13 &5.79 &-0.04 &83 &32.64 &0.0970 &5 &1.21 \\
40 &21.67 &8.13 &0.12 &77 &37.28 &20.32 &8.76 &0.08 &78 &37.12 &0.0091 &1 &0.16 \\ \hline
41 &10.55 &3.71 &0.20 &90 &25.57 &12.21 &5.73 &-0.15 &87 &26.30 &0.0473 &3 &0.73 \\
42 &12.29 &4.62 &0.05 &88 &28.57 &13.25 &5.50 &-0.01 &86 &29.03 &0.0068 &2 &0.47 \\
43 &19.54 &7.58 &0.08 &79 &34.55 &14.17 &5.92 &-0.22 &85 &33.39 &0.1475 &6 &1.16 \\
44 &9.91 &3.67 &0.04 &90 &27.56 &12.03 &5.61 &-0.29 &88 &28.16 &0.0527 &2 &0.60 \\
45 &27.46 &12.61 &-0.15 &70 &42.44 &18.73 &11.13 &-0.27 &79 &40.66 &0.1070 &9 &1.78 \\ \hline
46 &7.88 &4.47 &0.12 &92 &28.97 &12.10 &5.61 &0.08 &87 &30.73 &0.0745 &5 &1.76 \\
47 &14.85 &8.63 &0.12 &83 &33.64 &13.03 &4.85 &-0.02 &87 &32.62 &0.0743 &4 &1.03 \\
48 &11.07 &6.77 &-0.02 &88 &32.02 &16.23 &6.64 &-0.15 &83 &33.39 &0.0665 &5 &1.37 \\
49 &12.44 &6.28 &0.05 &87 &35.07 &14.00 &4.64 &0.01 &86 &35.32 &0.0399 &1 &0.26 \\
50 &10.05 &3.95 &0.11 &90 &31.20 &14.89 &5.19 &0.03 &85 &32.89 &0.1159 &5 &1.69 \\ \hline
\multicolumn{11}{r|}{\bf Avg.}& {\bf0.0702 } & {\bf 3.34} & {\bf 0.69} \\
\hline
\end{tabular}
}
\label{Table:second_jnd_result}
\end{table*}

\begin{table*}[!htbp]
\centering
\caption{GEV distribution model of the \textit{third} JND for the 50 image sequences of the MCL-JCI dataset. Shown are the location $\mu$, scale $\sigma$ and shape $\xi$, for both ground truth and SUR-FeatNet, together with the 50\% JND values and PSNR at the 50\% JND value. The Bhattacharyya distance measures the divergence between the predicted and ground truth distributions, $\Delta \text{JND} = |\widehat{\text{JND}} - \text{JND}|$, and  $\Delta \text{PSNR} = |\widehat{\text{PSNR}} - \text{PSNR}|$. 
}
\resizebox{0.99\textwidth}{!}{
\begin{tabular}{c|rrrrr|rrrrr|c|r|r}
\hline
Image & \multicolumn{5}{c|}{Ground truth}   & \multicolumn{5}{c|}{SUR-FeatNet}   & Bhattacharyya & \multicolumn{1}{c|}{$\Delta \text{JND}$} & \multicolumn{1}{c}{$\Delta \text{PSNR}$}\\
$k$ & \multicolumn{1}{c}{$\mu$} & \multicolumn{1}{c}{$\sigma$} & \multicolumn{1}{c}{$\xi$} & \multicolumn{1}{c}{\text{JND}} & \multicolumn{1}{c|}{\text{PSNR}} & \multicolumn{1}{c}{$\hat{\mu}$} & \multicolumn{1}{c}{$\hat{\sigma}$} & \multicolumn{1}{c}{$\hat{\xi}$} & \multicolumn{1}{c}{$\widehat{\text{JND}}$} & \multicolumn{1}{c|}{$\widehat{\text{PSNR}}$} & distance & & (dB)\\
\hline
1 &11.56 &3.36 &-0.35 &89 &29.18 &9.91 &5.63 &0.09 &89 &29.18 &0.1480 &0 &0.00 \\
2 &17.34 &7.63 &0.09 &81 &37.95 &13.23 &12.34 &0.39 &83 &37.62 &0.1180 &2 &0.33 \\
3 &10.61 &4.93 &0.04 &89 &29.18 &9.44 &4.25 &0.08 &90 &28.89 &0.0077 &1 &0.28 \\
4 &11.56 &4.60 &-0.06 &88 &27.04 &11.30 &5.18 &0.16 &88 &27.04 &0.0162 &0 &0.00 \\
5 &13.48 &5.39 &0.10 &86 &29.79 &12.06 &6.01 &0.26 &87 &29.53 &0.0182 &1 &0.25 \\ \hline
6 &11.94 &4.17 &-0.02 &88 &30.81 &10.54 &4.65 &-0.02 &89 &30.50 &0.0182 &1 &0.31 \\
7 &16.66 &10.38 &0.01 &81 &27.95 &9.74 &5.88 &-0.02 &90 &26.34 &0.1286 &9 &1.61 \\
8 &14.31 &4.23 &-0.08 &86 &26.79 &12.32 &5.91 &-0.03 &87 &26.59 &0.0531 &1 &0.20 \\
9 &8.61 &2.66 &0.20 &92 &25.67 &10.07 &5.17 &-0.15 &90 &26.30 &0.0804 &2 &0.63 \\
10 &10.59 &3.82 &0.05 &89 &33.98 &10.08 &4.88 &0.22 &90 &33.60 &0.0245 &1 &0.38 \\ \hline
11 &16.01 &6.21 &-0.10 &83 &32.48 &11.76 &4.45 &-0.07 &88 &31.55 &0.0823 &5 &0.93 \\
12 &14.95 &9.33 &0.34 &83 &30.57 &13.72 &7.49 &0.21 &85 &30.23 &0.0167 &2 &0.34 \\
13 &10.34 &3.04 &-0.08 &90 &32.43 &10.52 &5.13 &-0.00 &89 &32.84 &0.0675 &1 &0.41 \\
14 &8.32 &3.27 &0.28 &92 &29.86 &10.06 &4.79 &0.00 &90 &30.54 &0.0354 &2 &0.67 \\
15 &9.13 &2.56 &0.24 &91 &24.67 &10.74 &3.59 &0.20 &89 &25.04 &0.0298 &2 &0.37 \\ \hline
16 &10.40 &3.99 &0.06 &90 &28.32 &10.68 &4.51 &0.02 &89 &28.58 &0.0028 &1 &0.26 \\
17 &10.26 &2.74 &-0.05 &90 &27.49 &9.40 &3.14 &0.02 &91 &27.24 &0.0190 &1 &0.25 \\
18 &11.24 &3.89 &-0.39 &89 &31.19 &9.57 &6.32 &-0.20 &90 &30.85 &0.1113 &1 &0.34 \\
19 &13.23 &5.19 &-0.38 &86 &28.09 &13.48 &5.95 &0.10 &86 &28.09 &0.0852 &0 &0.00 \\
20 &17.46 &8.83 &-0.21 &81 &30.04 &12.63 &9.44 &-0.18 &86 &29.19 &0.0487 &5 &0.85 \\ \hline
21 &15.69 &7.66 &-0.08 &83 &27.10 &13.28 &6.76 &-0.05 &86 &26.57 &0.0147 &3 &0.53 \\
22 &9.32 &2.74 &0.20 &91 &26.41 &9.78 &4.75 &0.24 &90 &26.68 &0.0528 &1 &0.27 \\
23 &11.37 &4.23 &-0.10 &89 &25.10 &11.86 &5.86 &0.01 &87 &25.48 &0.0317 &2 &0.38 \\
24 &12.73 &4.76 &-0.16 &87 &31.07 &11.03 &5.09 &-0.01 &89 &30.53 &0.0201 &2 &0.55 \\
25 &9.85 &2.97 &0.18 &91 &26.19 &10.34 &4.83 &0.16 &89 &26.75 &0.0418 &2 &0.56 \\ \hline
26 &15.16 &7.93 &-0.00 &83 &31.06 &15.63 &7.90 &0.07 &83 &31.06 &0.0032 &0 &0.00 \\
27 &8.69 &2.01 &0.21 &92 &26.88 &8.78 &3.17 &-0.05 &92 &26.88 &0.0491 &0 &0.00 \\
28 &11.40 &5.15 &-0.10 &88 &36.58 &13.27 &4.67 &0.06 &86 &37.27 &0.0346 &2 &0.68 \\
29 &13.38 &4.49 &-0.45 &87 &34.18 &9.57 &5.00 &0.02 &90 &33.19 &0.1082 &3 &0.99 \\
30 &11.24 &4.58 &-0.32 &89 &32.80 &10.01 &3.85 &0.15 &90 &32.39 &0.0527 &1 &0.41 \\ \hline
31 &8.91 &3.49 &0.26 &91 &30.03 &9.31 &4.35 &0.07 &91 &30.03 &0.0144 &0 &0.00 \\
32 &9.99 &4.21 &-0.01 &90 &28.13 &8.49 &3.51 &-0.11 &92 &27.36 &0.0325 &2 &0.77 \\
33 &9.31 &3.13 &0.23 &91 &29.59 &11.65 &6.30 &-0.07 &88 &30.57 &0.0835 &3 &0.99 \\
34 &12.25 &4.66 &0.20 &87 &29.13 &10.70 &4.17 &0.03 &89 &28.66 &0.0270 &2 &0.46 \\
35 &9.71 &3.78 &-0.01 &90 &28.32 &9.40 &4.17 &0.06 &91 &27.99 &0.0052 &1 &0.33 \\ \hline
36 &12.58 &4.51 &-0.11 &87 &28.35 &11.71 &4.52 &0.01 &88 &28.13 &0.0071 &1 &0.22 \\
37 &18.62 &13.16 &0.09 &78 &26.83 &13.64 &7.66 &0.27 &85 &25.78 &0.0548 &7 &1.05 \\
38 &8.14 &3.00 &0.35 &92 &26.72 &10.53 &5.01 &0.01 &89 &27.64 &0.0585 &3 &0.92 \\
39 &10.50 &5.88 &0.09 &89 &31.12 &12.15 &6.15 &0.08 &87 &31.72 &0.0093 &2 &0.61 \\
40 &16.20 &6.83 &0.19 &83 &36.36 &16.19 &6.07 &0.17 &83 &36.36 &0.0043 &0 &0.00 \\ \hline
41 &9.62 &2.48 &0.26 &91 &25.28 &9.38 &4.23 &-0.21 &91 &25.28 &0.0892 &0 &0.00 \\
42 &10.17 &4.27 &-0.20 &90 &27.99 &11.19 &4.67 &0.15 &89 &28.28 &0.0530 &1 &0.29 \\
43 &14.69 &6.81 &-0.05 &84 &33.61 &10.34 &4.38 &-0.08 &90 &31.93 &0.1005 &6 &1.68 \\
44 &8.53 &2.56 &-0.05 &92 &26.85 &10.37 &4.11 &0.01 &90 &27.56 &0.0721 &2 &0.71 \\
45 &17.90 &10.94 &-0.06 &80 &40.40 &13.29 &7.33 &-0.08 &86 &38.36 &0.0651 &6 &2.04 \\ \hline
46 &8.65 &2.98 &0.18 &92 &28.97 &9.33 &4.87 &0.04 &90 &29.77 &0.0421 &2 &0.81 \\
47 &10.53 &6.63 &0.18 &88 &32.31 &9.70 &4.32 &-0.07 &90 &31.61 &0.0589 &2 &0.69 \\
48 &8.27 &4.37 &0.26 &92 &30.36 &11.51 &4.91 &0.00 &88 &32.02 &0.0537 &4 &1.66 \\
49 &8.84 &3.15 &0.49 &91 &33.71 &10.89 &3.65 &-0.04 &89 &34.46 &0.0642 &2 &0.75 \\
50 &8.46 &2.92 &0.25 &92 &30.26 &13.32 &4.06 &0.08 &87 &32.30 &0.1956 &5 &2.04 \\ \hline
\multicolumn{11}{r|}{\bf Avg.}& {\bf 0.0522 } & {\bf 2.10} & {\bf 0.58} \\
\hline
\end{tabular}
}
\label{Table:third_jnd_result}
\end{table*}

\begin{table*}[!htbp]
\centering
\caption{GEV distribution model of the \textit{first} JND for the 40 image sequences of the JND-Pano dataset. Shown are the location $\mu$, scale $\sigma$ and shape $\xi$, for both ground truth and SUR-FeatNet, together with the 50\% JND values and PSNR at the 50\% JND value. The Bhattacharyya distance measures the divergence between the predicted and ground truth distributions, $\Delta \text{JND} = |\widehat{\text{JND}} - \text{JND}|$, and  $\Delta \text{PSNR} = |\widehat{\text{PSNR}} - \text{PSNR}|$. 
}
\resizebox{0.99\textwidth}{!}{
\begin{tabular}{c|rrrrr|rrrrr|c|r|r}
\hline
Image & \multicolumn{5}{c|}{Ground truth}   & \multicolumn{5}{c|}{SUR-FeatNet}   & Bhattacharyya & \multicolumn{1}{c|}{$\Delta \text{JND}$} & \multicolumn{1}{c}{$\Delta \text{PSNR}$}\\
$k$ & \multicolumn{1}{c}{$\mu$} & \multicolumn{1}{c}{$\sigma$} & \multicolumn{1}{c}{$\xi$} & \multicolumn{1}{c}{\text{JND}} & \multicolumn{1}{c|}{\text{PSNR}} & \multicolumn{1}{c}{$\hat{\mu}$} & \multicolumn{1}{c}{$\hat{\sigma}$} & \multicolumn{1}{c}{$\hat{\xi}$} & \multicolumn{1}{c}{$\widehat{\text{JND}}$} & \multicolumn{1}{c|}{$\widehat{\text{PSNR}}$} & distance & & (dB)\\
\hline
1 &29.63 &12.90 &0.09 &67 &33.24 &33.03 &13.43 &0.12 &63 &33.70 &0.0242 &4 &0.45 \\
2 &35.18 &13.03 &0.01 &62 &41.09 &32.88 &12.01 &-0.25 &64 &40.80 &0.0388 &2 &0.28 \\
3 &26.04 &6.80 &0.53 &73 &31.82 &30.50 &11.53 &-0.05 &67 &32.38 &0.0845 &6 &0.57 \\
4 &32.72 &12.13 &-0.29 &65 &30.17 &29.74 &11.23 &0.09 &68 &29.88 &0.0398 &3 &0.29 \\
5 &33.24 &13.77 &-0.07 &63 &33.14 &24.68 &13.39 &-0.08 &72 &32.11 &0.0490 &9 &1.03 \\ \hline
6 &38.12 &14.34 &-0.27 &58 &34.92 &24.10 &18.08 &-0.24 &71 &33.71 &0.0957 &13 &1.21 \\
7 &30.44 &10.74 &0.04 &67 &29.68 &42.34 &17.77 &-0.20 &53 &31.02 &0.0963 &14 &1.34 \\
8 &38.62 &14.03 &-0.48 &58 &30.15 &35.43 &13.55 &0.01 &61 &29.88 &0.0699 &3 &0.28 \\
9 &47.82 &17.18 &-0.37 &48 &31.47 &36.41 &15.02 &-0.19 &60 &30.53 &0.0561 &12 &0.95 \\
10 &51.36 &19.06 &-0.36 &44 &39.58 &36.44 &10.64 &-0.06 &61 &38.43 &0.1538 &17 &1.15 \\ \hline
11 &36.30 &10.14 &-0.43 &62 &34.78 &33.93 &13.02 &0.04 &63 &34.71 &0.0987 &1 &0.07 \\
12 &24.46 &12.11 &-0.51 &73 &32.08 &39.15 &14.50 &-0.02 &57 &33.62 &0.3741 &16 &1.54 \\
13 &39.70 &10.69 &-0.65 &58 &37.59 &30.45 &11.57 &0.13 &67 &36.76 &0.1650 &9 &0.84 \\
14 &29.51 &10.30 &-0.19 &68 &34.02 &36.38 &12.77 &0.04 &60 &34.69 &0.0870 &8 &0.67 \\
15 &36.14 &9.92 &-0.32 &62 &27.82 &30.00 &9.96 &0.06 &68 &27.34 &0.0636 &6 &0.48 \\ \hline
16 &40.28 &11.89 &-0.23 &57 &32.53 &34.52 &11.52 &-0.28 &63 &32.04 &0.0400 &6 &0.49 \\
17 &37.81 &13.43 &-0.18 &59 &30.96 &35.72 &16.66 &-0.18 &60 &30.92 &0.0152 &1 &0.04 \\
18 &60.23 &19.32 &-0.53 &35 &37.59 &49.24 &19.92 &-0.46 &46 &36.63 &0.0522 &11 &0.96 \\
19 &44.24 &16.07 &-0.37 &52 &31.98 &32.15 &11.80 &0.14 &65 &30.83 &0.0990 &13 &1.15 \\
20 &44.08 &15.01 &-0.26 &52 &33.16 &35.47 &18.08 &-0.35 &60 &32.48 &0.0461 &8 &0.68 \\ \hline
21 &39.22 &17.56 &-0.20 &56 &30.26 &43.01 &17.59 &-0.34 &52 &30.63 &0.0135 &4 &0.37 \\
22 &30.85 &17.16 &-1.13 &66 &29.79 &49.28 &19.62 &-0.31 &45 &31.26 &0.2440 &21 &1.47 \\
23 &38.91 &18.06 &-0.44 &56 &28.97 &24.44 &15.53 &-0.25 &72 &27.49 &0.1070 &16 &1.48 \\
24 &24.38 &12.24 &-0.26 &73 &33.27 &33.25 &14.62 &-0.08 &63 &34.22 &0.1023 &10 &0.95 \\
25 &36.19 &13.41 &-0.64 &61 &30.67 &34.12 &11.81 &0.07 &63 &30.52 &0.1217 &2 &0.15 \\ \hline
26 &49.02 &25.93 &-0.47 &44 &35.29 &37.45 &11.58 &-0.22 &60 &33.93 &0.2197 &16 &1.36 \\
27 &35.24 &11.90 &0.06 &62 &32.01 &32.21 &11.08 &-0.20 &65 &31.74 &0.0396 &3 &0.27 \\
28 &42.54 &6.89 &-0.51 &57 &41.89 &40.02 &16.29 &-0.24 &56 &41.95 &0.2577 &1 &0.06 \\
29 &40.22 &20.17 &-0.29 &54 &37.83 &40.32 &13.36 &-0.20 &56 &37.72 &0.0382 &2 &0.11 \\
30 &34.49 &14.66 &-0.09 &62 &37.89 &37.41 &16.04 &-0.10 &58 &38.22 &0.0101 &4 &0.33 \\ \hline
31 &42.14 &15.15 &-0.29 &54 &35.92 &34.64 &13.29 &-0.20 &62 &35.25 &0.0335 &8 &0.67 \\
32 &26.39 &5.64 &0.93 &73 &31.55 &31.06 &12.40 &-0.17 &66 &32.38 &0.1928 &7 &0.83 \\
33 &41.31 &15.01 &-0.30 &55 &35.20 &34.87 &12.59 &-0.28 &62 &34.60 &0.0444 &7 &0.60 \\
34 &38.78 &12.73 &-1.04 &59 &32.65 &42.39 &20.40 &-0.20 &52 &33.24 &0.2886 &7 &0.59 \\
35 &24.82 &10.13 &0.05 &73 &31.61 &42.45 &14.44 &-0.26 &54 &33.54 &0.1920 &19 &1.93 \\ \hline
36 &57.31 &19.34 &-0.59 &38 &33.25 &34.85 &13.87 &-0.20 &62 &31.45 &0.2220 &24 &1.80 \\
37 &35.29 &15.13 &-0.46 &61 &28.81 &38.70 &13.54 &-0.13 &58 &29.13 &0.0630 &3 &0.32 \\
38 &34.65 &13.24 &-0.22 &62 &31.54 &41.20 &16.99 &-0.42 &55 &32.17 &0.0361 &7 &0.63 \\
39 &58.57 &23.59 &-1.05 &36 &38.21 &41.03 &19.27 &-0.48 &54 &36.42 &0.1706 &18 &1.79 \\
40 &40.85 &18.90 &-0.47 &54 &39.28 &45.57 &16.59 &-0.14 &50 &39.52 &0.0651 &4 &0.24 \\ \hline
\multicolumn{11}{r|}{\bf Avg.}& {\bf 0.1053 } & {\bf 8.63} & {\bf 0.76} \\
\hline
\end{tabular}
}
\label{Table:pano_first_jnd_result}
\end{table*}

\end{document}